\documentclass[preprint,12pt,authoryear]{elsarticle}

\usepackage{amssymb}
\usepackage{natbib}
\usepackage{subfigure,graphicx,psfrag,epsf}
\usepackage{amsmath,amsfonts,latexsym,amssymb,euscript,xr}
\usepackage{multirow}
\usepackage{dsfont}
\usepackage{pdflscape}
\usepackage{tikz}
\usetikzlibrary{bayesnet}
\usepackage{bigints}
\usepackage{algorithm}
\usepackage{algpseudocode}
\usepackage{multirow}

\journal{--}

\begin{document}

\begin{frontmatter}



\title{Bayesian Smooth-and-Match strategy for ordinary differential equations models that are linear in the parameters}


\author[BO]{Saverio Ranciati \corref{cor}}\ead{saverio.ranciati2@unibo.it}
\author[BO]{Cinzia Viroli}
\author[GR]{Ernst Wit}
\cortext[cor]{corresponding author.}

\address[BO]{Department of Statistical Sciences, University of Bologna, Via Belle Arti 41, Bologna, Italy}
\address[GR]{University of Groningen, Johann Bernoulli Institute of Mathematics and Computer Science, Nijenborgh 9, Groningen, the Netherlands}

\begin{abstract}
In many fields of application, dynamic processes that evolve through time are well described by systems of ordinary differential equations (ODEs). The analytical solution of the ODEs is often not available and different methods have been proposed to infer these quantities: from numerical optimization to regularized (penalized) models, these procedures aim to estimate indirectly the parameters without solving the system. We focus on the class of techniques that use smoothing to avoid direct integration and, in particular, on a Bayesian Smooth-and-Match strategy that allows to obtain the ODEs' solution while performing inference on models that are linear in the parameters. We incorporate in the strategy two main sources of uncertainty: the noise level in the measurements and the model error. We assess the performance of the proposed approach in three different simulation studies and we compare the results on a dataset on neuron electrical activity.
\end{abstract}


\end{frontmatter}


\section{Introduction}

Many processes that evolve through time are described by systems of ordinary differential equations (ODEs). Imagine we have a simple case of one ODE where the change, $dx(t)$, of the concentration level of a specific molecule in the cell follows a law that is described by some function $g_{\beta}[x(t)]$, with a set of parameters $\boldsymbol{\beta}$ governing this law. We could think of this change as depending on the quantity of molecules at time $t$, that is $x(t)$, a rate $\beta_1$ at which new ones are produced by the cell as time passes by and then maybe some `limit' $\beta_2$ on the capacity to contain them. What we observe, in practice, are not the actual changes $dx(t)$ (the derivative of the process) but instead the concentration levels $x(t)$ at sampling time points. This means that, in order to relate the data at our disposal to the parameters of interest, we need a solution for the system of differential equations: however, in most of the cases, no closed forms are available and numerical techniques are thus needed. Moreover, our observations may very well be affected by noise that perturbs the true temporal dynamic of the process. There are several techniques in the literature on this topic (see \citet{robinson2004introduction} for an introduction): most of them involve numerical integration, a straightforward approach to the problem that, however, does not take into account in any way the uncertainty about the chosen statistical model nor the noise level in the data. Also, methods relying on this type of solvers require an explicit computation of the solution at every step of the algorithm, severely hindering the procedure in practice. A way to avoid direct numerical integration (or differentiation) is \emph{smoothing} the data. An example of a \emph{one-step} kernel-based nonparametric smooth estimator was recently proposed in \citet{RSSB:RSSB12040}. In general, the idea of smoothing to avoid integration falls under the class of \emph{collocation} methods, in which some of them are called \emph{two-steps} \citep{liang2012parameter,gugushvili2012n,dattner2013estimation}. As in \citet{varah1982spline,madar2003incorporating,brunel2008parameter}, a first step consist of recovering a temporary solution of the system by smoothing or interpolating the data (i.e. with cubic splines, least squares, nonparametric filters, local polynomial regression and so forth) and then applying nonlinear least squares to infer the parameters of the ODEs. The properties of these methods, such as consistency and asymptotic normality, are discussed in \citet{xue2010sieve}. Other methods following a similar strategy of smoothing and matching are discussed in \citet{gonzalez2013inferring,campbell2012smooth}.
Another approach is to use regularization \citep{Ramsay,gonzalez2014reproducing,Vujacic}, in order to do inference on the parameters while minimizing - with a frequentist flavor - some measure of distance between the theoretical solution and the estimated one. An initial guess of the parameters of interest is provided to the algorithm, as it is used together with a linear combination of basis functions to solve a penalized optimization problem.
From a Bayesian point of view, as in \citet{chkrebtii2013bayesian}, Gaussian Processes (GP) are prominent tools employed to solve the task. They encode naturally a source of randomness in the solution and simultaneously provide a class of flexible priors for the functions used to smooth the data coming from the ODEs' system. A recent approach using GP is provided by \citet{calderhead2009accelerating}, with some drawbacks that were later addressed in \citet{dondelinger2013ode} through the use of adaptive gradient matching. Another advantage of these Bayesian methods is the complete probabilistic phrasing of the problem, allowing for a statistical quantification of the uncertainty about the solution obtained: the core of these procedures are - in fact - probabilistic solvers that can be sampled to explore the parameter space while obtaining indirectly a solution of the system \citep{conrad2015probability}.  For some other (implicit and explicit) probabilistic solvers see \citet{barber2014solving}; some applications of these methods on real life dataset are presented in \citet{honkela2010model} and \citet{titsias2012identifying}.

In this work, we propose a two-step Bayesian strategy (Bayesian Smooth-and-Match) that borrows the idea of smoothing to overcome direct integration and, simultaneously, to filter some of the noise in the data. The first step of the method relies on penalized splines to smooth the data and reconstruct the variables of the ODEs; the second step focuses on inferring the parameters of the system through ridge regression, with covariates being known functions of the process that is being studied.

The rest of the manuscript is organized as follows: in Section \ref{mod_sec}, the notation used to build the strategy is introduced and we discuss the distributional assumptions on the data, together with the prior and posterior distributions; in Section \ref{sim_sec}, numerical and visual results on three different simulation studies are reported; in Section \ref{nagumo_sec}, an application to a previously analyzed dataset is presented; finally, in Section \ref{conc_sec}, we provide a summary of the work and we outline some future developments.

\section{Model formulation}\label{mod_sec}
\subsection{Tools and notation}

Suppose we observe a $p$-dimensional vector $\boldsymbol{y}_i$ containing noisy observations of a dynamic process $\boldsymbol{x}(t_i)$, where $t_i$ is a generic element from the set of ordered time points $\left\{ t_i \right\} \in [0,1]$, and $i=1,\dots,n$ indexing the independent observed vectors. For the sake of simplicity, we will drop the ``$(t)$'' argument  in the notation for $\boldsymbol{x}(t)$ and its components when not ambiguous, otherwise it will be explicit. We assume that:
\begin{equation}
\label{lik1}
y_{ik} \sim N\left( x_{ik}, \sigma^2_k \right)
\end{equation}
with $k=1, \dots, p$, $x_{ik}=x_k(t_i)$ and $\sigma^2_k$ a parameter describing the noise level in the data for the component $k$ of the whole process $\boldsymbol{x}$. For every observation $t_i$, each component $\boldsymbol{x}_{\cdot k}$ can be approximated by a cubic spline \citep[p. 122]{Wood} of $t$ in the time domain
\begin{equation}
\label{def_x}
x_{ik} = \sum_{h=1}^{q_k+2} \theta_{hk} \psi_{hk}  (t_i)
\end{equation}
where the sum is over $q_k+2$ known basis functions $\psi_{hk} (\cdot)$ that depend on the $q_k$ number of arbitrary knots used to construct them. For an arbitrary fixed $t$, we assume that the dynamic process $\boldsymbol{x}(t)$ is well described by an ordinary differential equations (ODEs) model defined as:
\begin{equation}
\label{system}
\begin{cases}
\boldsymbol{x}'(t)= \frac{d\boldsymbol{x}(t)}{dt} = g(\boldsymbol{x}(t)) \\ \boldsymbol{x}(0)= \boldsymbol{\xi}
\end{cases}
\end{equation}
where $\boldsymbol{x}'(t)$ is the first derivative with respect to time of a continuous process $\boldsymbol{x}(t)=(\boldsymbol{x}_{\cdot 1}(t), \dots, \boldsymbol{x}_{\cdot k}(t), \dots, \boldsymbol{x}_{\cdot p}(t))$, $\boldsymbol{\xi}=(\xi_1,\dots,\xi_k,\dots,\xi_p)$ is a vector of initial conditions for the system and $g : \mathds{R}^p \rightarrow \mathds{R}^p$ a (possibly) non-linear function of $\boldsymbol{x}(t)$. The generic functional form of the derivative for the first variable $\boldsymbol{x}_{\cdot 1}$ is
\begin{equation}
\label{g_ode}
\frac{d\boldsymbol{x}_{\cdot 1}(t)}{dt}=g_1(\boldsymbol{x}(t))= \sum_{j=1}^b \beta_j h_j(\boldsymbol{x}(t))
\end{equation}
which involves the first element  $g_1 : \mathds{R}^p \rightarrow \mathds{R}$ of the function $g$. We think of it as a linear combination of $b$ parameters of interest and some functions $h_j$, with $j=1,\dots,b$, that describe the dynamic evolution of the component $\boldsymbol{x}_{\cdot 1}$. Instead of working with the derivative of $\boldsymbol{x}$ we switch to the integral representation of the system in Equation \ref{system}:
\begin{eqnarray}
\label{sol_def}
\boldsymbol{x}_{i1}=\int_{t_1}^{t_i}  \frac{d\boldsymbol{x}_{\cdot 1}(t)}{dt}  &=& \int_{t_1}^{t_i} g_1(\boldsymbol{x}(s))ds = \\
\nonumber &=& \xi_1 + \int_{t_1}^{t_i} \sum_{j=1}^b \beta_j h_j(\boldsymbol{x}(s)) ds
\end{eqnarray}
for $i=2,\dots,n$, because we set our first observation $t_1$ as the starting time point ($t_1$ could be either zero or not), and $\xi_1=x_{11}$. The initial condition $\xi_1$ can be either estimated or assumed known. The solution for the first ODE, given by Equation \ref{sol_def}, is thus:
\begin{equation}
\label{g1}
\boldsymbol{x}_{i1} = \xi_1 +  \sum_{j=1}^b \left( \beta_j  \int_{t_1}^{t_i} h_j(\boldsymbol{x}(s)) ds \right)= \xi_1 + \sum_{j=1}^b \beta_j H_j (\boldsymbol{x})
\end{equation}
where $H_j$ is obtained by numerical integration of the corresponding function $h_j$ from $t_1$ to $t_i$. As it will be clarified in the next section, we have now two distinct definitions (Equation \ref{def_x} and Equation \ref{g1}) for the same quantity $\boldsymbol{x}_{\cdot 1}$: this redundant representation gives rise to two different distributional assumptions on the corresponding first observed vector $\boldsymbol{y}_{\cdot 1}$. We focus on ODE models that are linear in the parameters so that the problem can be later rephrased in a regression framework.

\subsection{Prior, likelihood and posterior distributions}

With reference to a sample of $n$ observations, Equation (\ref{lik1}) may be rewritten as:
$$ \boldsymbol{y}_{\cdot k}=(y_{1k}, \dots,y_{nk})^{T} \sim \mathcal{N}_n \left( \boldsymbol{x}_{\cdot k}, \sigma^2_k \mbox{I}_n \right)$$
with $\sigma^2_k$ the noise level for component $k$ and $\mbox{I}_n$ the identity matrix of order $n$. From Equation \ref{def_x}, $\boldsymbol{x}_{\cdot k}=\Psi_k \boldsymbol{\theta}_k$ where $\Psi$ is the $n \times (q_k+2)$ matrix of spline basis evaluated at every time point $t_i$. Thus, assuming that every component is independent from the other given the column vectors $\boldsymbol{\theta}_k$, the likelihood function of the model is
$$ \label{joint_link} P(\boldsymbol{y} | \boldsymbol{\Theta}, \boldsymbol{\sigma}^2) = \prod_{i=1}^{n} \prod_{k=1}^{p} P(y_{ik} | \boldsymbol{\theta}_k, \sigma^2_k)$$
where $\boldsymbol{y}=(\boldsymbol{y}_{\cdot 1},\dots,\boldsymbol{y}_{\cdot p})$, $\boldsymbol{\Theta}=\{ \boldsymbol{\theta}_1,\dots,\boldsymbol{\theta}_p \}$ and $\boldsymbol{\sigma}^2=(\sigma^2_1,\dots,\sigma^2_p)$. We choose to tackle the inferential procedure with a Bayesian approach: this allows us to represent the whole process with a fully probabilistic generative model that we can also describe as a graphical model. Furthermore, within the Bayesian framework, we can take into account the variability and the uncertainty at every level of the problem, from the smoothing of the data to the estimation of the parameters governing the system, incorporating the two main sources of noise: the measurement error and the model error. The former will be quantified by the variances $\sigma^2_k$, for $k=2,\dots,p$; with $\sigma^2_1$ will bring together both sources of error (model and measurement) into the same parameter. We assign prior probabilities to the parameters of the splines: we assume, for each $\boldsymbol{\theta}_k$, a Gaussian prior distribution of the form
$$ \boldsymbol{\theta}_k |\, \lambda_{\theta_k} \sim \mathcal{N}_{q_k+2} \left(\boldsymbol{0}, \left[ \lambda_{\theta_k} \mbox{S}_{\theta_k} \right]^{-1} \right).$$
Choosing this prior is equivalent to performing a penalized spline regression on the data: the form of the hyperparameter precision matrix $\mbox{S}_{\theta_k}$ defines the way we want the basis to be penalized. The elements on the first two rows and columns of $\mbox{S}_{\theta_k}$ are zeros while, for $l=3,\dots,q_k+2,l'=3,\dots,q_k+2$, the generic element $(l,l')$ is equal to $\psi_{ll'}$. We follow this approach to ensure that non-linearity in the components is captured without, however, producing curves that would overfit the noisy data. The parameter $\lambda_{\theta_k}$ penalizes the non-smoothness of the functions $\psi_{hk}$: the `wiggliness' of the curve resulting from the spline smoothing is encoded in the precision matrix $\mbox{S}_{\theta_k}$ (see \citet{Wood} for further details, p. 126). These penalization terms have prior distributions
$$ \lambda_{\theta_k} |\, \alpha_{\theta_k},\gamma_{\theta_k} \sim \mbox{Gam}(\alpha_{\theta_k},\gamma_{\theta_k})$$
for some vectors of shape and rate hyperparameters $(\boldsymbol{\alpha}_{\theta}, \boldsymbol{\gamma}_{\theta})$. These two-dimensional vectors are chosen to represent weakly informative priors on $\lambda_{\theta_k}$: more specifically, we select values of the hyperparameters that encourage undersmoothing of the data \citep{gugushvili2012n}, with enough variance for the Gamma distribution to be able to shift to a more penalized curve if needed. A reference improper prior density $P(\sigma^2_k)=1/\sigma^2_k$ is employed for each $\sigma^2_k$. The first vector of observations $\boldsymbol{y}_{\cdot 1}$ has another representation, stemming directly from the ODEs system's solution for $\boldsymbol{x}_{\cdot 1}$ that depends on the integral solution from Equation \ref{g1}, that is
$$ \boldsymbol{y}^\star_{\cdot 1}|\, \xi_1, \boldsymbol{\beta}, \boldsymbol{\Theta}, \sigma^2_1 \sim \mathcal{N}_n \left( \xi_1 \boldsymbol{1}_n + \boldsymbol{H} \boldsymbol{\beta} , \sigma^2_1 \mbox{I}_n \right)$$
where $\boldsymbol{1}_n$ is a $n$-dimensional unitary column vector, $\boldsymbol{\beta}$ the column vector of parameters we want to estimate and $\boldsymbol{H}$ a $n \times b$ matrix collecting the integrated functions $H_j$. As in our case, when the starting inverse-problem may be `ill-posed', sure enough the ordinary least squares estimation leads to overdetermined or underdetermined systems of equations as solution to the regression itself. Regularization is the usual approach to overcome this issue and  Tikhonov regularization, in particular, is one of the most commonly used: from a Bayesian point of view, it is equivalent to assume, for $\boldsymbol{\beta}$, the following prior distribution
$$ \boldsymbol{\beta} |  \lambda_{\beta} \sim \mathcal{N}_{b}\left( \boldsymbol{0}, [\lambda_{\beta} \mbox{I}_b]^{-1} \right).$$
The choice of this prior, effectively, induces the Bayesian `ridge regression' with $\lambda_{\beta}$ acting as a penalizing term for which we assume a prior distribution
$$ \lambda_{\beta}| \alpha_{\beta}, \gamma_{\beta} \sim \mbox{Gam}(\alpha_{\beta}, \gamma_{\beta}).$$
As for the other penalization terms, we follow the same approach of using a weakly informative prior. We also estimate $\xi_1$ instead of rescaling the data and we select a flat prior for it.
Dropping hyperparameters from the notation, the joint posterior distribution of the parameters in our model is:
\begin{eqnarray}
\label{joint}
\nonumber P(\boldsymbol{\Theta}, & \xi_1 &, \boldsymbol{\beta}, \boldsymbol{\lambda}_{\theta}, \lambda_{\beta}, \boldsymbol{\sigma}^2 | \, \boldsymbol{y}, \boldsymbol{y}^\star_{\cdot 1} ) \propto P\!\left( \boldsymbol{y}^\star_{\cdot 1}|\, \xi_1, \boldsymbol{\beta}, \boldsymbol{\Theta}, \sigma^2_1 \right) \times \\
\nonumber & \times & \prod^p_{k=1} \left[ P\!\left( \boldsymbol{y}_{\cdot k} |\, \boldsymbol{\theta}_k, \sigma^2_k \right) P\!\left(  \boldsymbol{\theta}_k |\, \lambda_{\theta_k} \right) P(\lambda_{\theta_k}) P(\sigma^2_k) \right] \times \\
& \times & P\!\left(  \boldsymbol{\beta} |\, \lambda_{\beta} \right)  P\!\left( \lambda_{\beta} \right) P(\xi_1)
\end{eqnarray}
which is represented in the graphical model in Figure \ref{graph_mod}.

\subsection{Mimicking the data: relationship with other methods}\label{mimic}

\begin{figure}[h!]
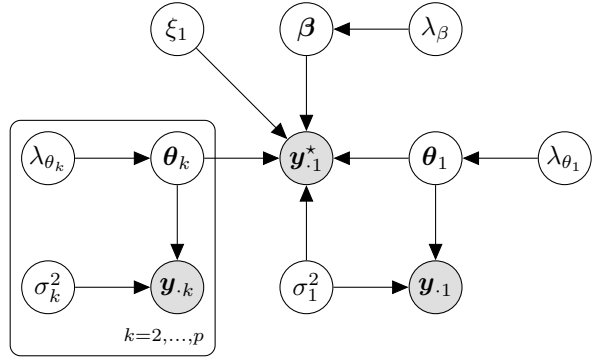

      \centering
      \tikz{ %
        \node[obs] (yk) {$\boldsymbol{y}_{\cdot k}$} ; %
        \node[latent, above=of yk] (theta) {$\boldsymbol{\theta}_k$} ; %
        \node[obs, right=of theta] (y1star) {$\boldsymbol{y}^\star_{\cdot 1}$} ; %
        \node[latent, above=of theta] (xi) {$\xi_1$};
        \node[latent, below=of y1star] (sigma1) {$\sigma^2_1$} ;
        \node[latent, above=of y1star] (beta) {$\boldsymbol{\beta}$} ; %
        \node[latent, left=of yk] (sigma) {$\sigma^2_k$} ;
        \node[latent, right=of beta] (lam_b) {$\lambda_{\beta}$} ;
        \node[latent, left=of theta] (lam_t) {$\lambda_{\theta_k}$} ;
        \node[obs, right=of sigma1] (y1) {$\boldsymbol{y}_{\cdot 1}$} ; %
	\node[latent, right=of y1star] (theta1) {$\boldsymbol{\theta}_1$} ; %
	\node[latent, right=of theta1] (lam_t1) {$\lambda_{\theta_1}$} ;
	\edge {theta} {y1star} ; %
	\edge {beta} {y1star} ;
	\edge {sigma1} {y1star} ;
	\edge {lam_b} {beta} ;
	\edge {sigma} {yk} ;
	\edge {theta} {yk} ;
	\edge {lam_t} {theta} ;
	\edge {xi} {y1star} ;
	\edge {sigma1} {y1} ; 
	\edge {lam_t1} {theta1} ;
	\edge {theta1} {y1} ;
	\edge {theta1} {y1star} ;
	\plate [yshift=0cm, xshift=0cm] {model} {(yk) (theta) (lam_t) (sigma)} {$\,\,\,\,\,\,\,\,\,\,\,\,\,\scriptstyle{\,k=2,\dots,p}$}; %
      }
      \caption{Graphical representation of the ODEs and solution model}
      \label{graph_mod}
    \end{figure}

One feature of this representation is that the vector of observations $\boldsymbol{y}_{\cdot 1} $ appears twice: as a term of the likelihood component, when $k=1$ in the first product, and then as the noisy solution of the ODEs' system, that is $\boldsymbol{y}^\star_{\cdot 1}$.  In \citet{icml2014c2_barber14}, their GPODE model focuses on a probabilistic generative model for the data and the graph representation contains two nodes for the same quantity (namely, the process itself). The authors model the system $\boldsymbol{x}(t)$ as coming from a Gaussian process (GP) and exploit the fact that differentiating the GP produces derivatives $\boldsymbol{x}'(t)$ that are still modeled as a Gaussian process with available analytical description of its kernel (see the original manuscript for more details). Then, they marginalize over the components $\boldsymbol{x}(t)$ with a standard convolution integral and model the data $\boldsymbol{y}$ with a Gaussian distribution; after that, they reintroduce $\boldsymbol{x}(t)$ and couple it with the obtained derivatives to measure the distance between the deterministic ODEs of the system and the ones estimated by the data. This approach faces some issues. As pointed out in \citet{macdonald2015controversy}, where the authors inspect the graph representation in \citet{icml2014c2_barber14}, having two nodes assigned to the same quantity is methodologically inconsistent. To solve the issue, they first introduce a dummy variable that mimics $\boldsymbol{x}(t)$, thus removing the inconsistency. However, the two nodes are still conceptually describing the very same quantity and a natural definition of this dependency would be an undirected edge between them: this addition, unfortunately, changes the graph from a directed acyclic graph to a chain graph which is not a probabilistic generative model anymore. To preserve its nature, they keep the two nodes separate from one another but highlight the consequences of this choice: when some of the noisy vectors $\boldsymbol{y}_{\cdot k}$ are not available (that is the case of partially observed systems), the model itself might be unidentifiable because of the likelihood not depending anymore on the parameters of the ODEs after marginalizing over the unobserved quantities. As we face the same issue with our proposed strategy, we are thus limited to situations where all the components of the process $\boldsymbol{x}(t)$ are observed (with noise). Given that a probabilistic definition of what could be the directed edge between $\boldsymbol{y}_{\cdot 1}$ and $\boldsymbol{y}^{\star}_{\cdot 1}$ (or vice versa) is not obvious, we couple the two quantities only by assuming they share the same variance $\sigma^2_1$ (as described in Figure \ref{graph_mod}). This shared nuisance parameter, however, suffers from the independence assumption on $\boldsymbol{y}_{\cdot 1}$ and $\boldsymbol{y}^{\star}_{\cdot 1}$. Suppose the simplest case where $\boldsymbol{y}^{\star}_{\cdot 1}=\boldsymbol{y}_{\cdot 1} \sim \mathcal{N}(\boldsymbol{0}, \sigma^2_1 I_n)$: having no edge between the two nodes is equivalent to $P(\boldsymbol{y}^{\star}_{\cdot 1},\boldsymbol{y}_{\cdot 1}|\sigma^2_1)=P(\boldsymbol{y}^{\star}_{\cdot 1}|\sigma^2_1)P(\boldsymbol{y}_{\cdot 1}|\sigma^2_1)$ which is roughly equal to a Normal distribution with variance parameter $\frac{\sigma^2_1}{2}$. Obviously, this would not happen if a directed edge were to be added to the graph, as the joint distribution of the vector and its copy would be instead $P(\boldsymbol{y}^{\star}_{\cdot 1},\boldsymbol{y}_{\cdot 1}|\sigma^2_1)=P(\boldsymbol{y}^{\star}_{\cdot 1}| \boldsymbol{y}_{\cdot 1}, \sigma^2_1)P(\boldsymbol{y}_{\cdot 1}|\sigma^2_1)= 1 \cdot P(\boldsymbol{y}_{\cdot 1}|\sigma^2_1)$, but we already discussed there is no clear way to describe a directed edge between the two nodes. This inconvenience is mostly present in a simulation environment. We start with a deterministic solution of an ODE system, $\boldsymbol{x}_{\cdot 1}$, and we perturb it with some noise $\tilde{\sigma}^2$ thus obtaining an observed vector $\boldsymbol{y}_{\cdot 1}$: recovering the noise level in the data is not equivalent anymore to estimating $\tilde{\sigma}^2$, as in our model we are bringing together in $\sigma^2_1$ two sources of noise - albeit coming from the very same vector of observations. In fact, $\sigma^2_1$ bears both the uncertainty about how good the solution of the set of ODEs and also how good is the smoothing of the first vector of data that we used as a regressor in solving indirectly the system. Such an undesired `mismatch' effect should not present itself as a problem when dealing with real world observations.

\subsection{Bayesian Smooth-and-Match}

Following the standard derivations for the full conditionals from Equation \ref{joint}, the samplers for $\boldsymbol{\theta}_k$ should consider the quantities
\begin{eqnarray} \label{post_tet} \nonumber P(\boldsymbol{\theta}_k | \sigma^2_k, \lambda_{\theta_k}, \boldsymbol{y}_{\cdot k}) & \propto & P\!\left( \boldsymbol{y}_{\cdot k} |\, \boldsymbol{\theta}_k, \sigma^2_k \right) P\!\left(  \boldsymbol{\theta}_k |\, \lambda_{\theta_k} \right) \times \\
& \times & P\!\left( \boldsymbol{y}^\star_{\cdot 1}|\, \xi_1, \boldsymbol{\beta}, \boldsymbol{\Theta}, \sigma^2_1 \right) \end{eqnarray}
which demands for a Metropolis-within-MCMC. After some mathematical steps, it is possible to show that, in the previous posterior distribution, the full conditionals of the parameters in $\boldsymbol{\Theta}$ are not in closed form. Our primary focus, however, is on the estimation of the $\boldsymbol{\beta}$ vector that contains the parameters truly describing the dynamic evolution of the ODE system. The spline smoothing step (Equation \ref{def_x}) is, instead, just a convenient approach to build the regression matrix $\boldsymbol{H}$ in Equation \ref{g1}. So, in order to have a more stable and faster MCMC scheme, we decide to aim for an approximation of the `true' posterior distribution for $\boldsymbol{\Theta}$ and we adopt the following Bayesian Smooth-and-Match strategy.
The procedure consists of two steps: 
\begin{itemize}
\item first (\emph{smooth} step), we do Bayesian penalized spline smoothing to recover the $\boldsymbol{x}_{\cdot k}$s through a Gibbs sampler that we obtain for $\boldsymbol{\theta}_k$ by temporarily not considering the integral solution term;
\item second (\emph{match} step), we plug-in the $\boldsymbol{x}_{\cdot k}$s, computed with the sampled values for $\boldsymbol{\theta}_k$ from the previous step, into the sampler for $\boldsymbol{\beta}$ through the matrix $\boldsymbol{H}$.
\end{itemize}
In practice, this allows us to drop the term $P\!\left( \boldsymbol{y}^{\star}_{\cdot 1}|\, \xi_1, \boldsymbol{\beta}, \boldsymbol{\Theta}, \sigma^2_1 \right)$ when sampling each $\boldsymbol{\theta}_k$. A more sophisticated smoothing could be performed (see \citet{morrissey2011inferring} for a state-of-art Bayesian spline regression) but we consider a simpler approach because we are only interested, at the first step, in recovering the components of the process to be used as regressors later at the second step of the procedure. Notice that, from a frequentist point of view, a consistent estimator of the ODEs' parameters can be obtained under mild conditions (namely, on the penalizing terms) when following this plug-in approach  \citep{gugushvili2012n}. The two steps of the strategy are not completely stand-alone compartments: as previously said, the quantity $\sigma^2_1$ connects both parts and acts as a measure for the uncertainty (from the smoothing and the regression) of the solution indirectly obtained by doing inference on $\boldsymbol{\beta}$.
The other parameters of the model are all updated with Gibbs samplers that follow from standard derivations for conjugated priors and likelihood terms (see Appendix). To avoid slow-mixing chains, a quick solution is to adopt a block-sampling implementation: in this case, $(\lambda_{\theta_k}, \boldsymbol{\theta}_k)$ are updated with $k$ independent Metropolis-Hastings (MH) steps, and a single MH step is performed for $(\lambda_{\beta}, \boldsymbol{\beta})$. In our simulations and analyzed dataset we had no such issue and we decided to opt for faster and simpler separate Gibbs samplers.

\section{Simulation studies}\label{sim_sec}

In this Section, we validate our proposed strategy by testing it with three different ODEs' systems, starting from a simple one component model (logistic growth) up to a three components epidemic model (HIV viral fitness). We simulate nine scenarios for each ODEs model, exploring three different level of noise and three sample sizes ($n=25,n=100,n=500$). The level of contamination of the data (low, medium or high), with Gaussian noise, is quantified through signal-to-noise ratio (SNR), that is the ratio between the standard deviation of the deterministic simulated solution (signal) and the standard deviation of the error term (noise) we use to perturb it. An increase in the strength of noise is equivalent to a decrease of the associated SNR, as the standard deviation of the signal at the numerator is fixed for a given sample size. For comparative purposes, we also inspect our Bayesian Smooth-and-Match strategy (\emph{SnM}) together with the collocation method implemented in the R package \emph{CollocInfer} \citep{Ramsay}; \emph{CollocInfer} is an adaptive gradient matching method based on spline smoothing. For each scenario, we run both algorithms on 100 independently simulated datasets and we summarize the results as the average across these replicates. The uncertainty about the estimated parameters is quantified for both algorithms through the mean square error (MSE), computed on the one hundred simulated samples. The tuning parameters for the collocation method, such as the number of knots, order of the basis for the splines and the penalization term, are selected with the functions provided in the package. The collocation method also needs initial guesses for the regression coefficients and we provide starting points drawn randomly from uniform distributions over ranges $[\boldsymbol{\beta}-4\boldsymbol{\beta};\boldsymbol{\beta}+4\boldsymbol{\beta}]$; when the lower bound of the range is negative for parameters that only exist as positive we set it to a small value, close to zero. Given the sensibility of \emph{CollocInfer} to the provided starting points, we control if convergence is achieved by the algorithm and we discard datasets that result in degenerate estimates for the parameters by checking the associated likelihood in the output. Those datasets are not considered when computing the averages and a measure of the number of actual datasets (NAD) used is provided. As for the visual representation of the results, Figures \ref{growth_fig} to \ref{hiv_fig} show the plot mosaics with the results for one randomly drawn dataset of the simulated one hundred.

\subsection{Logistic population growth}\label{log_sec}

\begin{landscape}
\begin{figure}
\begin{tabular}{lll}
\includegraphics[scale=0.3]{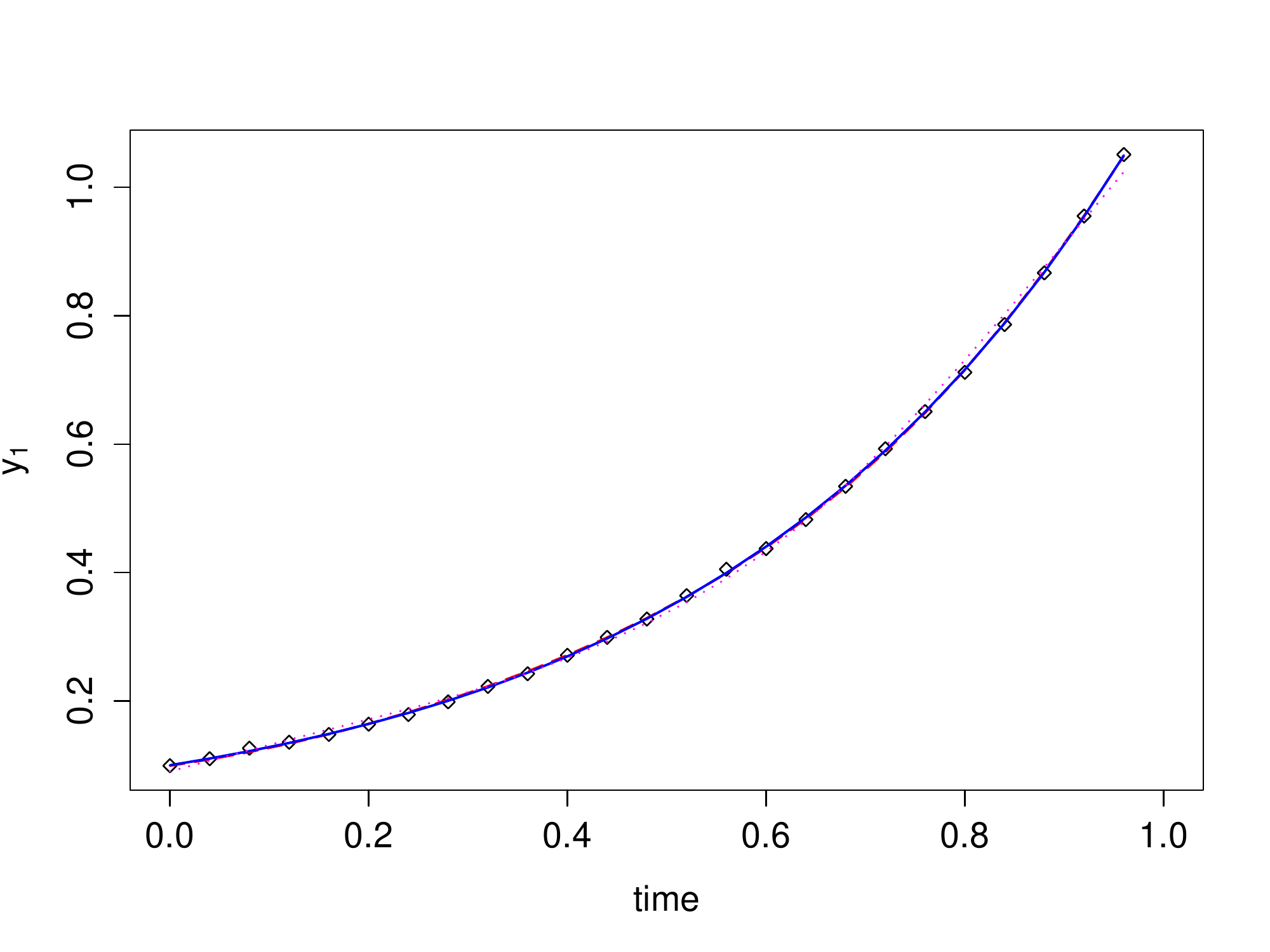} & \includegraphics[scale=0.3]{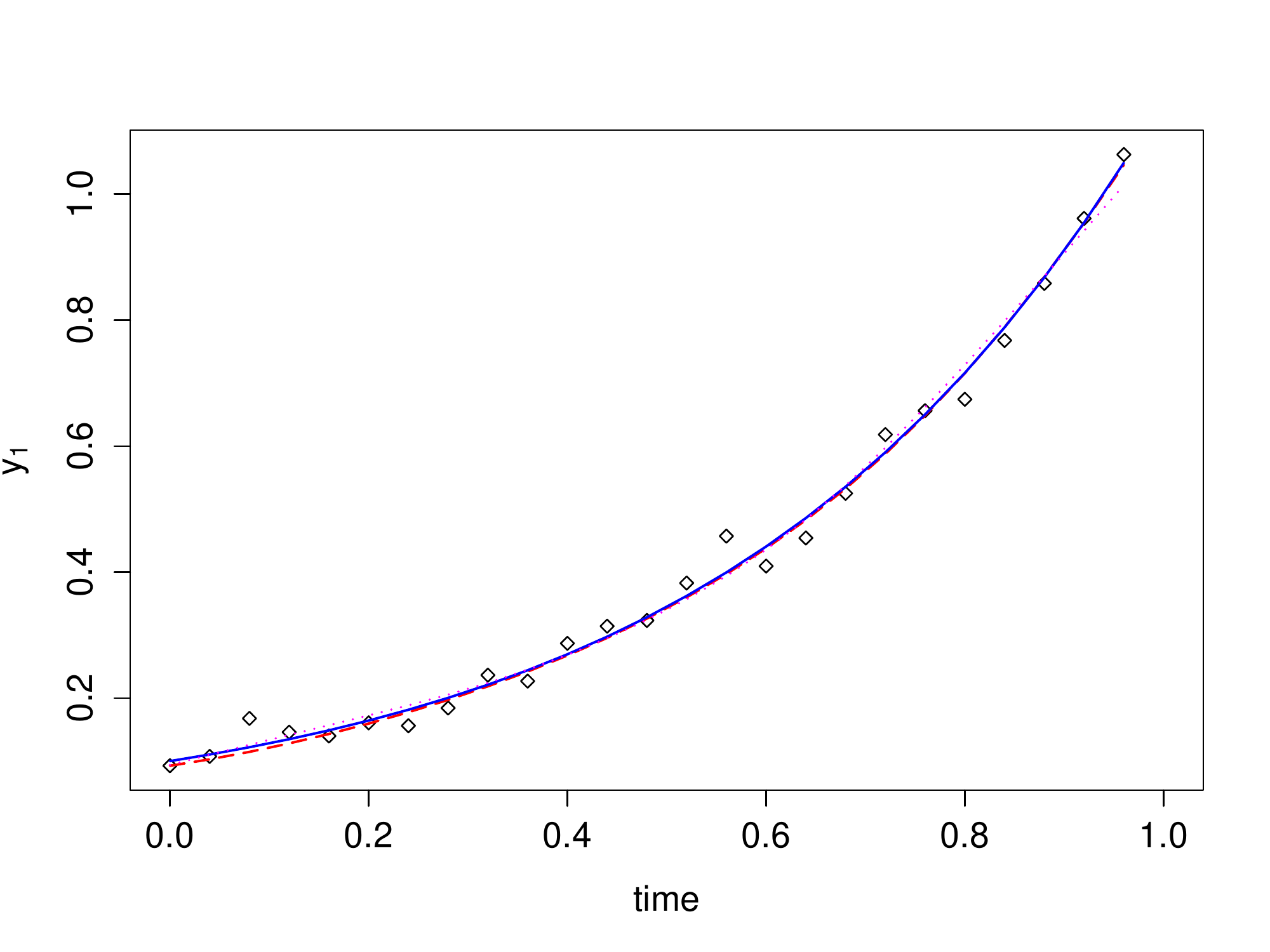} & \includegraphics[scale=0.3]{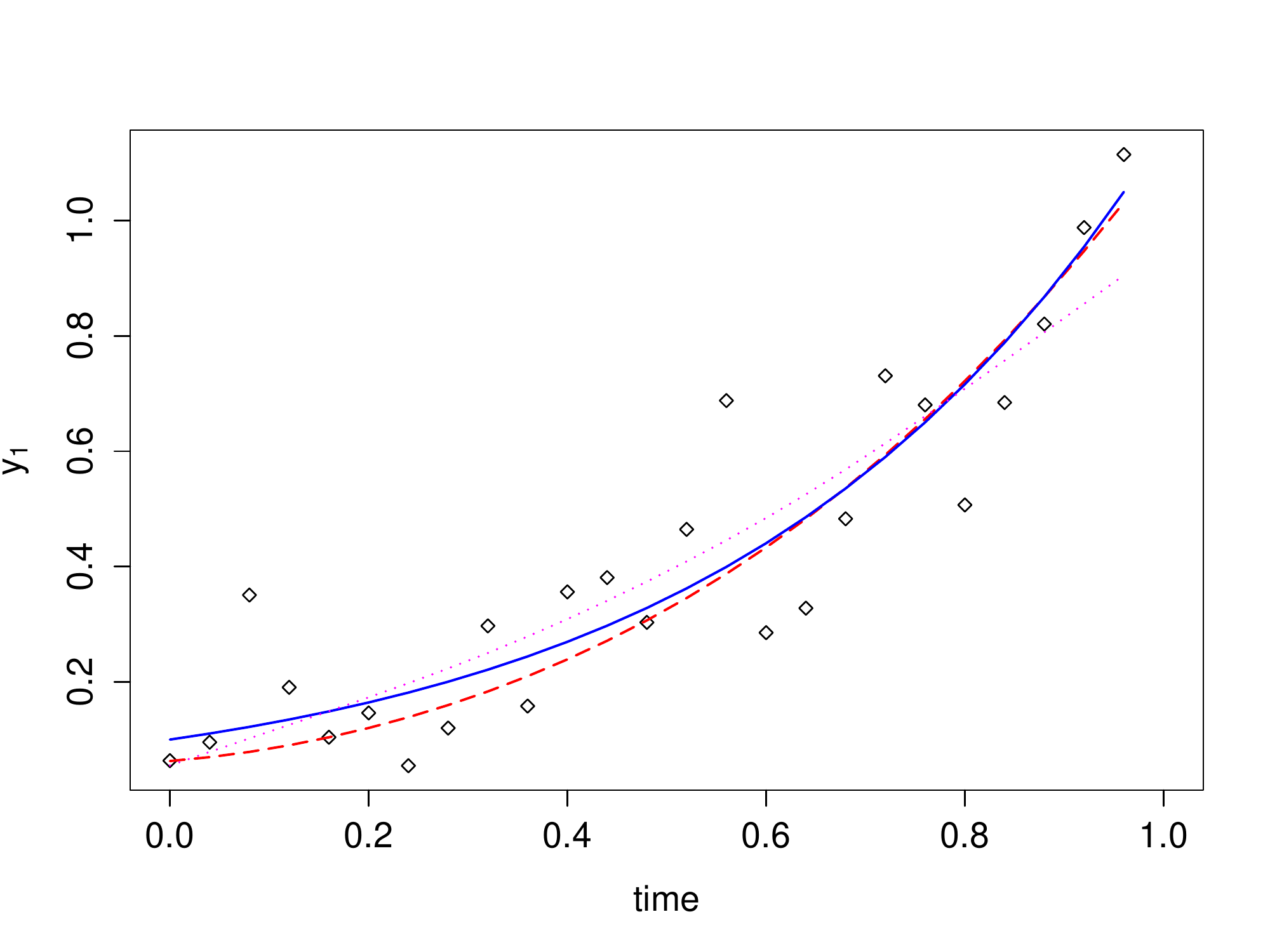} \\ 
\includegraphics[scale=0.3]{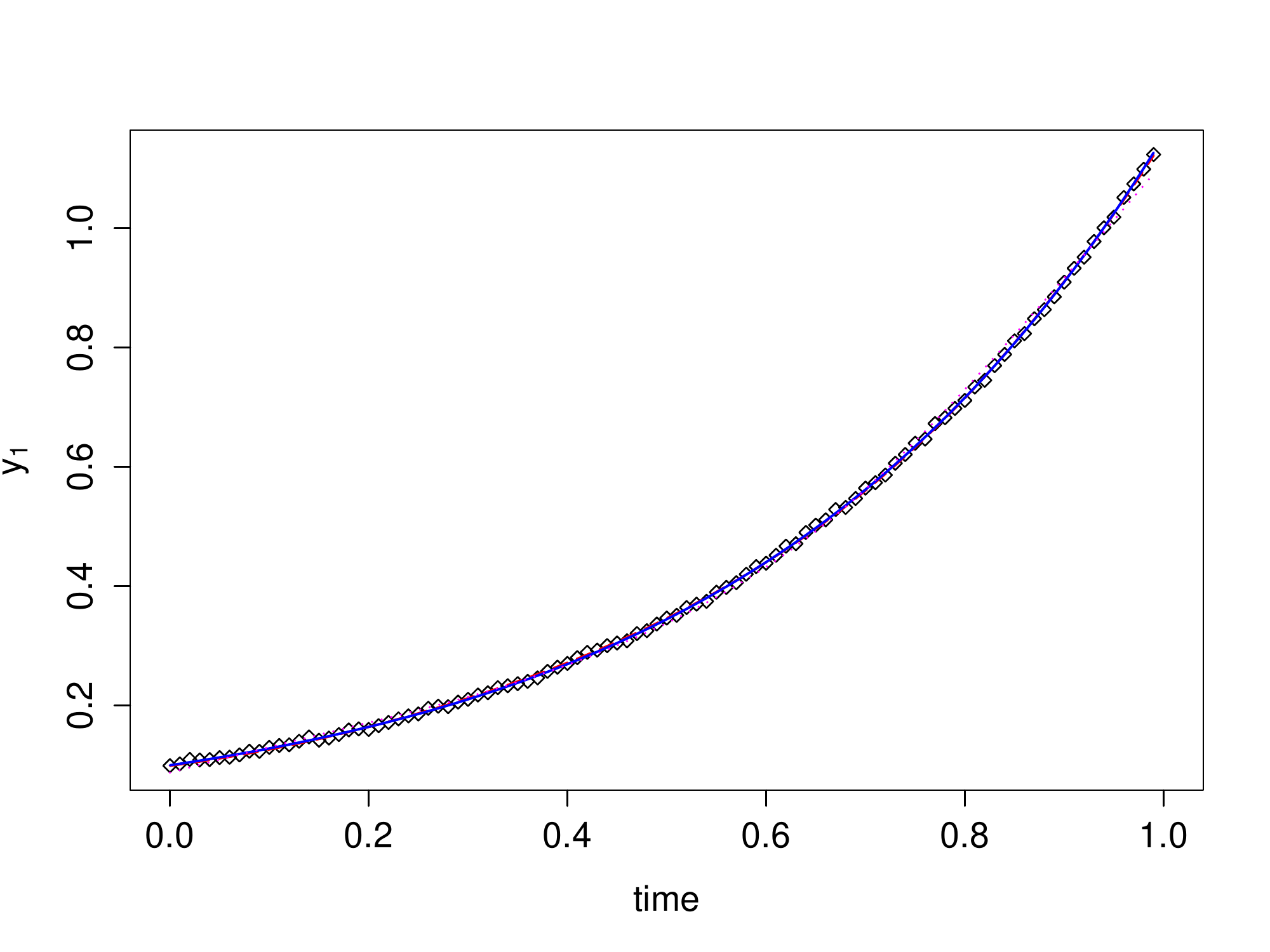} & \includegraphics[scale=0.3]{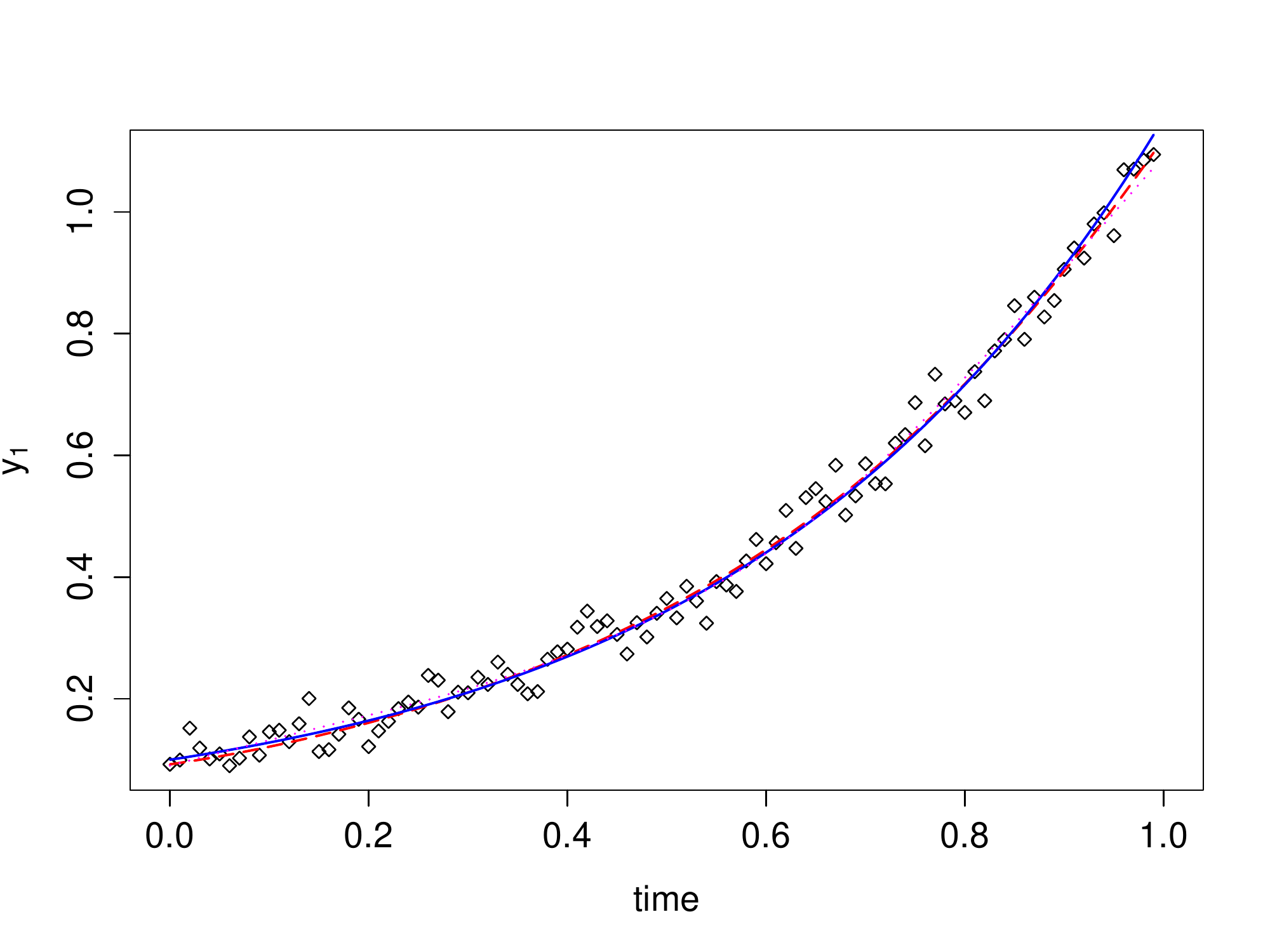} & \includegraphics[scale=0.3]{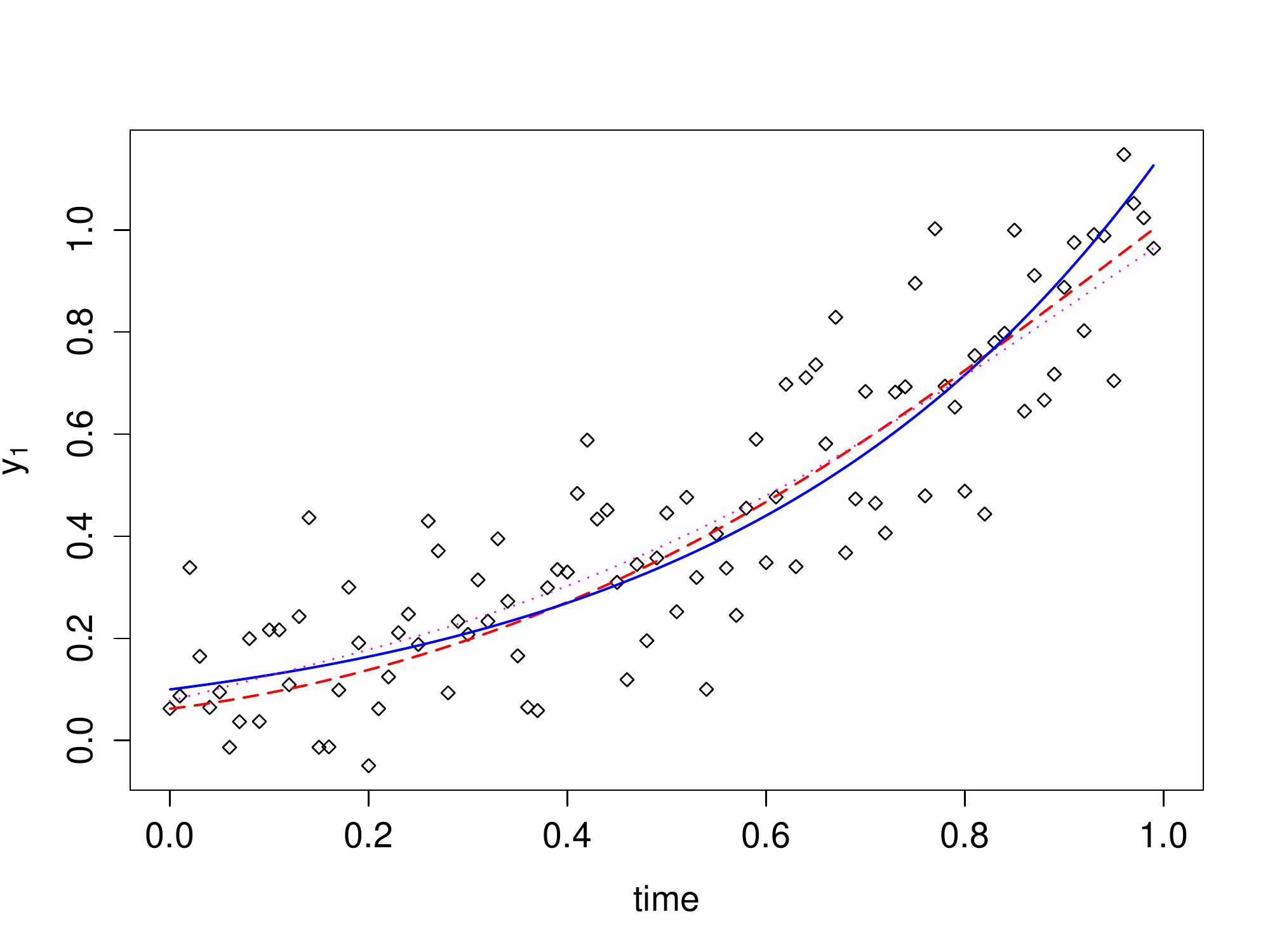} \\ 
\includegraphics[scale=0.3]{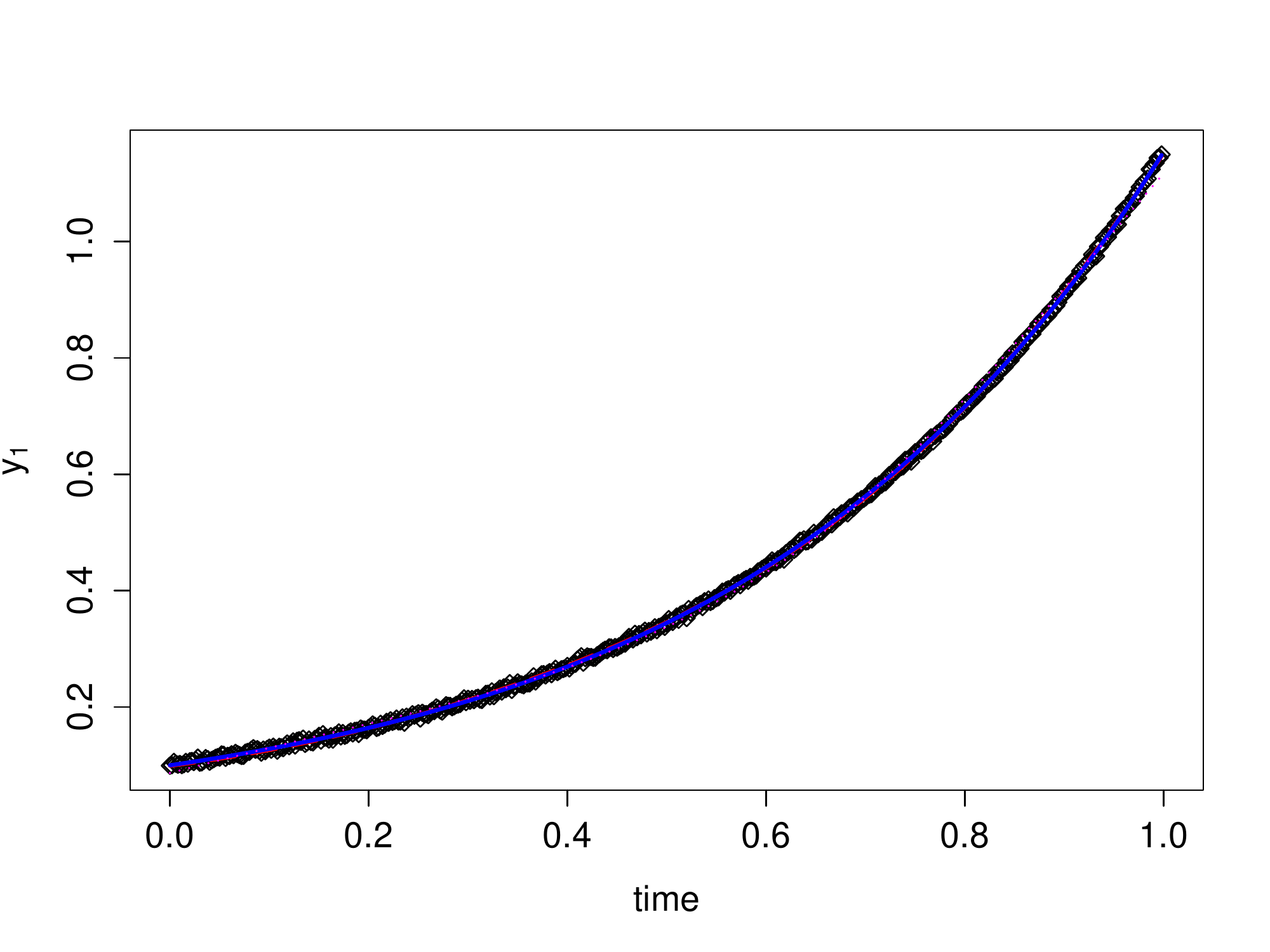} & \includegraphics[scale=0.3]{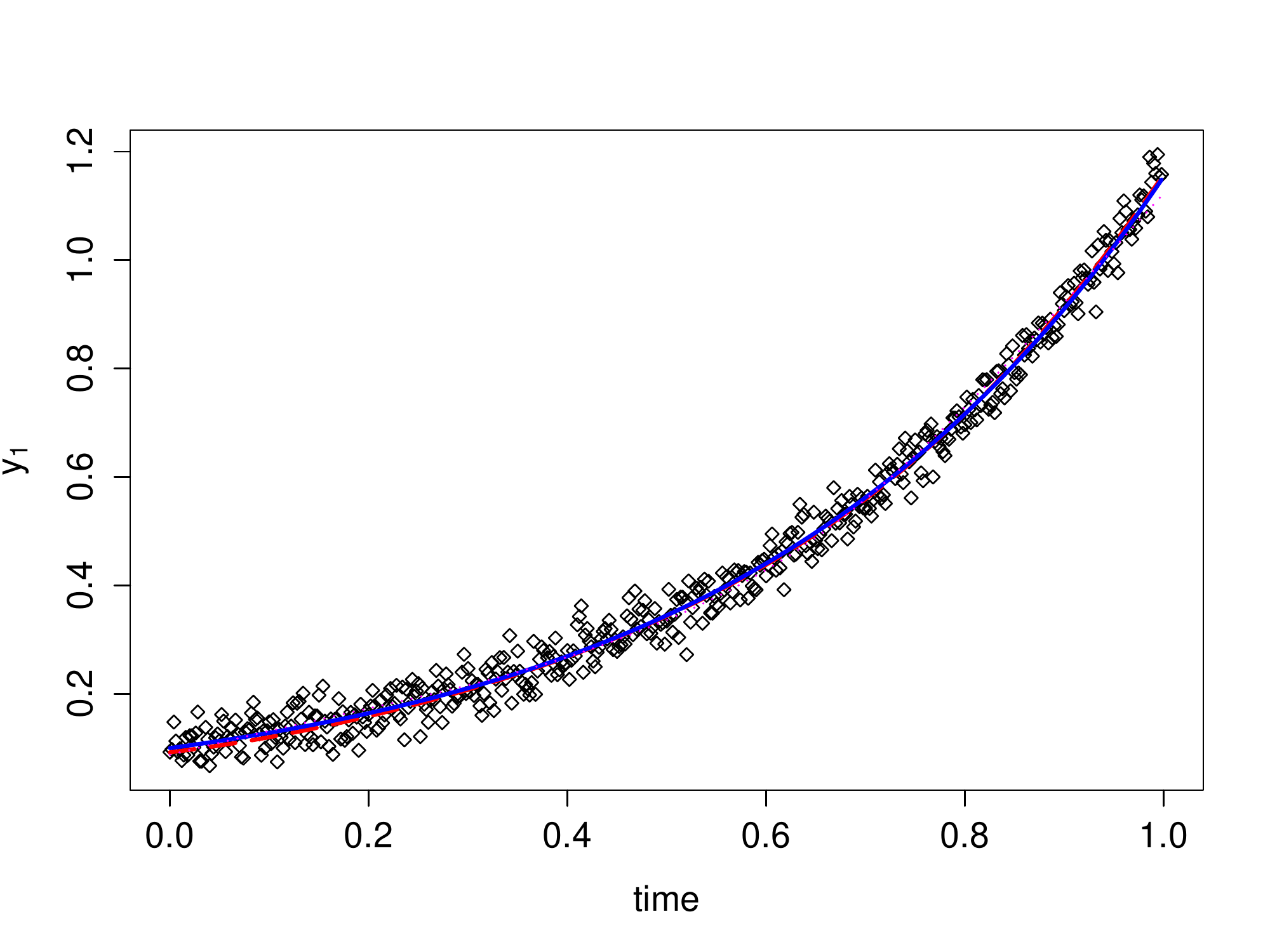} & \includegraphics[scale=0.3]{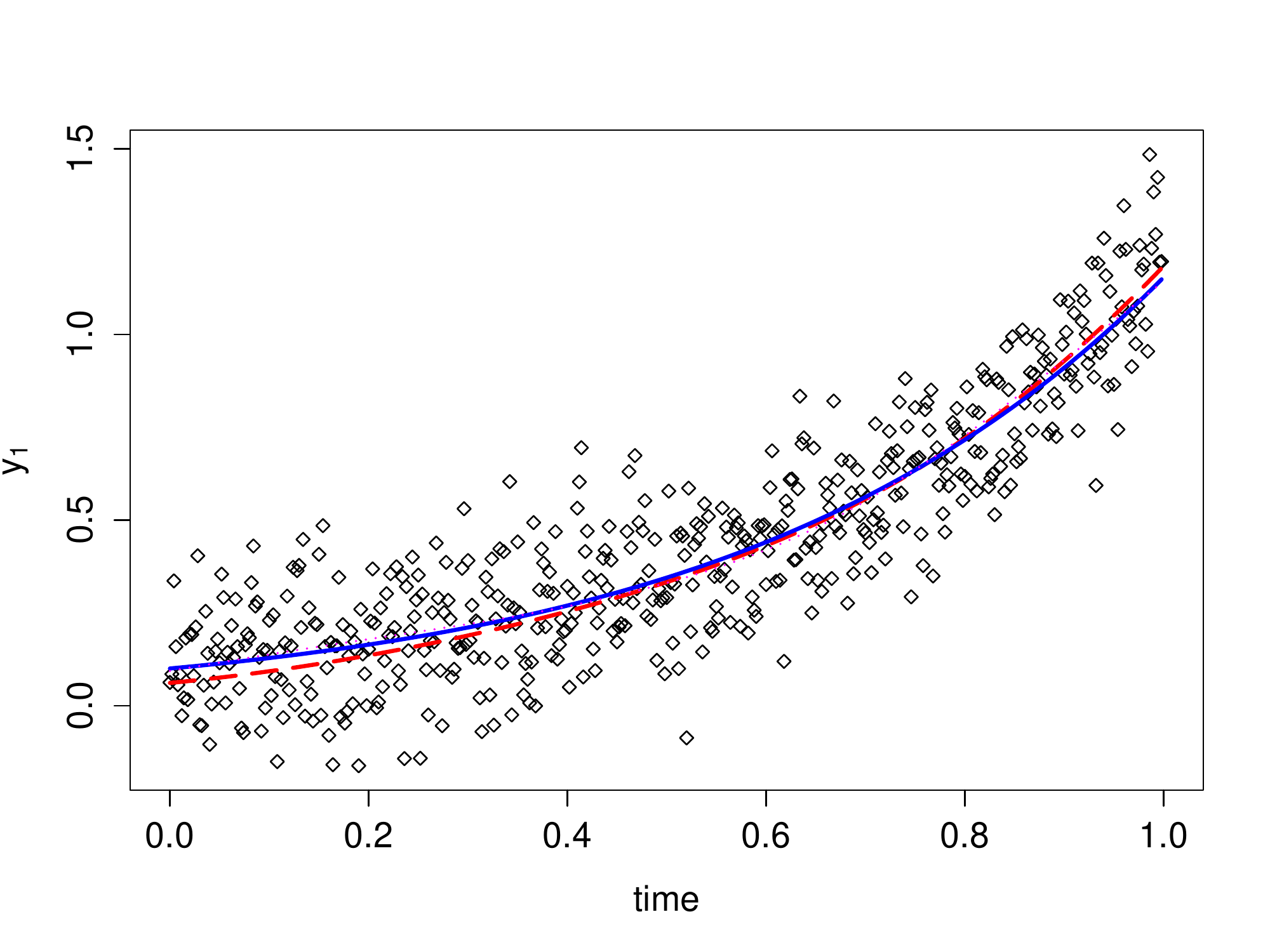} \\ 
\end{tabular}
\caption{Logistic population growth simulations: observed noisy data (\emph{dots}), smoothing spline (\emph{dotted} line), true solution of the ODE system (\emph{solid} line) and reconstructed solution (\emph{long-dashed} line) for the first variable $\boldsymbol{x}_{\cdot 1}$. Different sample sizes ($n=25,100,500$) from the top to the bottom and noise levels (low, medium and high) from left to right}\label{growth_fig}
\end{figure}
\end{landscape}

We first focus on a simple ODEs' system. We simulate data coming from the logistic growth model \citep{LogPopGrowth}, frequently employed in ecology and biology to describe the growth dynamics of a certain population. The system is defined as:
$$ \frac{d\boldsymbol{x}_{\cdot 1}}{dt}=a\boldsymbol{x}_{\cdot 1}\left(1-\frac{\boldsymbol{x}_{\cdot 1}}{K}\right),$$
where $a$ is the growth rate and $K$ the carrying capacity of the population involved. We consider another representation of previous equation, that is, in our notation,
$$\frac{d\boldsymbol{x}_{\cdot 1}}{dt}=\beta_1 \boldsymbol{x}_{\cdot 1} + \beta_2 \boldsymbol{x}^2_{\cdot 1},$$
linear in the parameters $(\beta_1=a,\beta_2=-a/K)$ we want to estimate, together with $x_{11}=\xi_1$. We simulate a noise-free solution of the system and then we perturb it with Gaussian errors at different level of noise (low, medium, high); the three levels are obtained by setting the standard deviation of the errors as increasing proportions of the mean value of $\boldsymbol{x}_{\cdot 1}$. The number of knots for the smoothing step, $q_1=2$, is manually selected, aiming for some undersmoothing of $\boldsymbol{y}_{\cdot 1}$ \citep{gugushvili2012n}. The functions used to build the regression matrix are $h_1(\boldsymbol{x})=\boldsymbol{x}_{\cdot 1}$ and $h_2(\boldsymbol{x})=\boldsymbol{x}_{\cdot 1}^2$; the true values for $\xi_1$, $\beta_1$ and $\beta_2$ used in the simulations are reported in Table \ref{growth_tab}. As expected, we can see from Table \ref{growth_tab} that to increasing levels of contamination of the data with noise correspond higher (on average) mean square errors for both methods. The average posterior mean of $\xi_1$ is stable throughout all the scenarios, showing some bias only when the SNR is at the highest level; \emph{CollocInfer} does not provide an estimate for the initial condition $\xi_1$. With \emph{SnM} we can recover the first parameter $\beta_1$ in almost any scenario with appreciable quality, also showing better results in comparison with \emph{CollocInfer}; when the noise contamination is at its maximum, however, the MSE computed on the one hundred posterior means is noticeably higher. The algorithm \emph{CollocInfer} seems to be more stable when retrieving the second parameter $\beta_2$, providing optimal estimates when the sample size is $n=100$ and the noise contamination up to a medium level ($SNR=13$ and $SNR=6.5$). The solution uncertainty quantified by \emph{SnM}, $\sigma^2_1$, appears to be less sensible to changes in sample size and more to the signal-to-noise ratio. In every scenario, the number of actual datasets (NAD) used to compute the results for \emph{CollocInfer} is less than 100, meaning that degenerate solutions were discarded in the process. A visual description of the results is presented in Figure \ref{growth_fig}: in most of the plots, the line describing the true curve (\emph{solid} line), the smoothed version of $\boldsymbol{y}_{\cdot 1}$ (\emph{dotted} line) and the ODE regression solution (\emph{long-dashed}) are undistinguishable from each other. They start to become appreciably different in the right part of the plots mosaic, showing the scenarios with the highest level of noise. We compute the average MSE between each curve (dotted line, $\mbox{MSE}_{x_1}$, and long-dashed line, $\mbox{MSE}_{g_1}$) and the true one representing the unperturbed data. For $n=500$ and $SNR=1.3$, the two mean square errors are $\mbox{MSE}_{x_1}=0.022$, for the smoothed reconstruction, and $\mbox{MSE}_{x_1}=0.007$ for the ODEs' system solution; with the same sample size but less noise, SNR=13, the difference between the two curves and the true one is almost negligible ($\mbox{MSE}_{x_1}=0.00015$ and $\mbox{MSE}_{g_1}=0.00016$).

\subsection{Lotka-Volterra} \label{lotka_sec}

\begin{landscape}
\begin{figure}
\begin{tabular}{lll}
\includegraphics[scale=0.3]{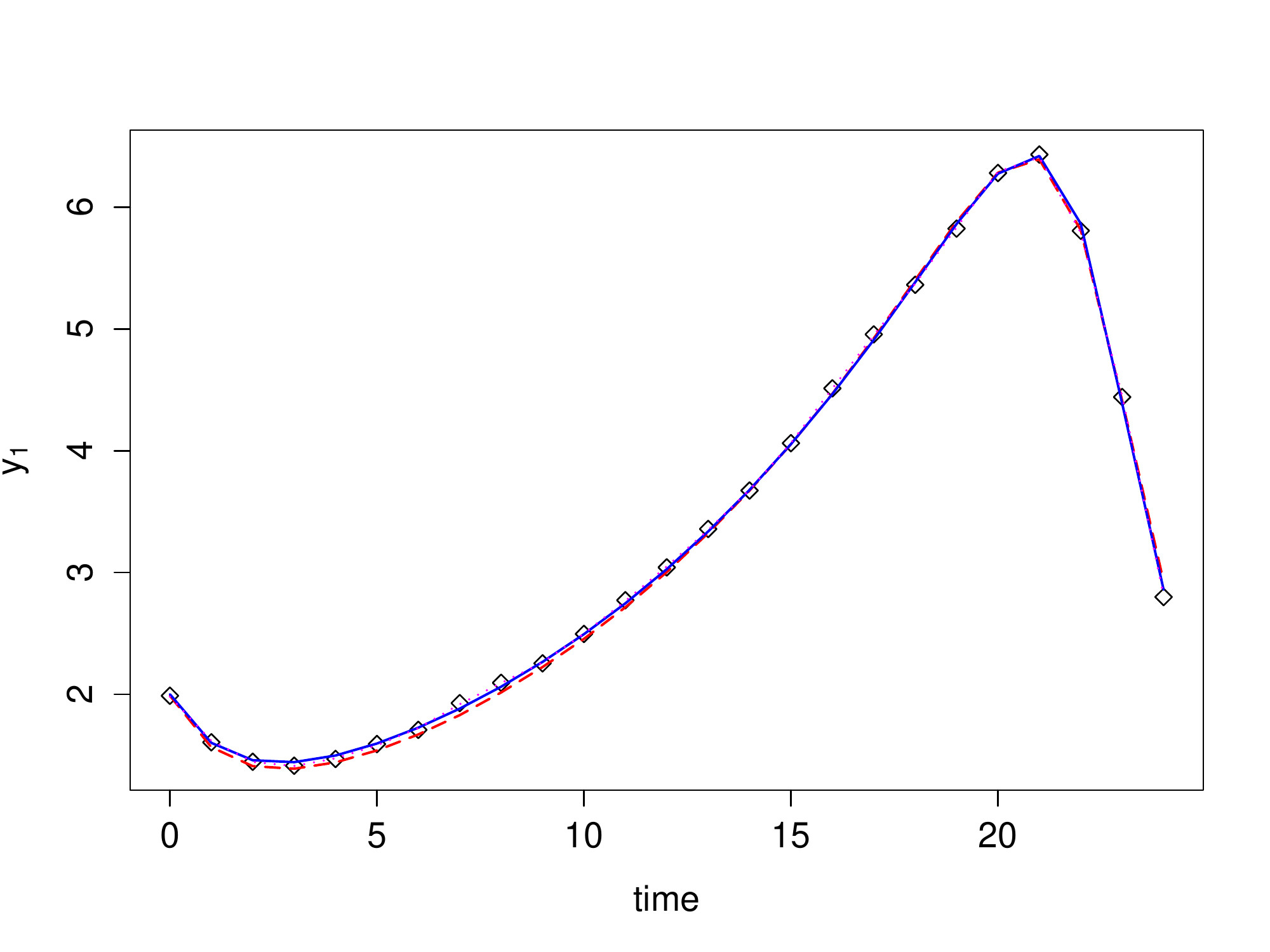} & \includegraphics[scale=0.3]{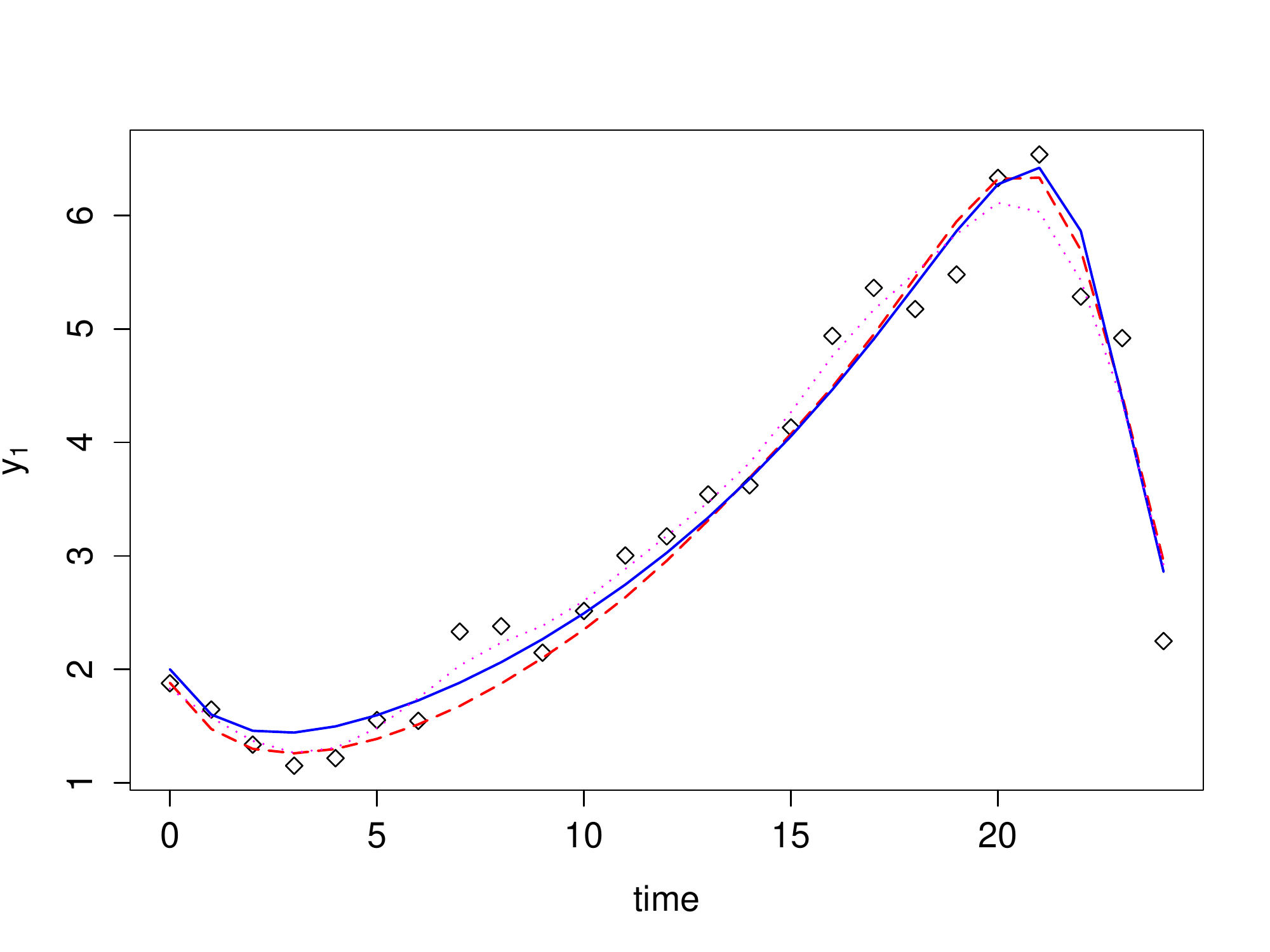} & \includegraphics[scale=0.3]{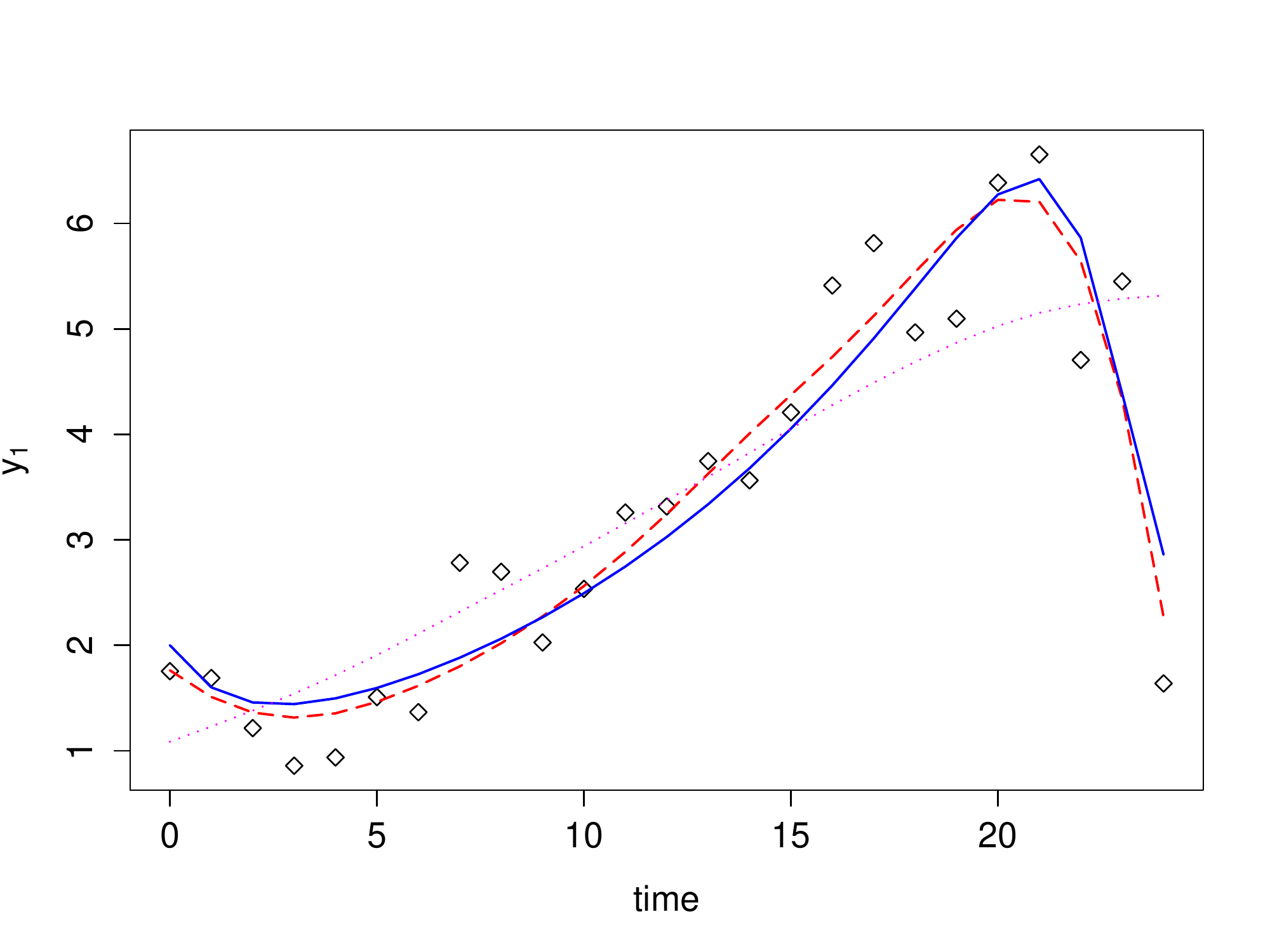} \\ 
\includegraphics[scale=0.3]{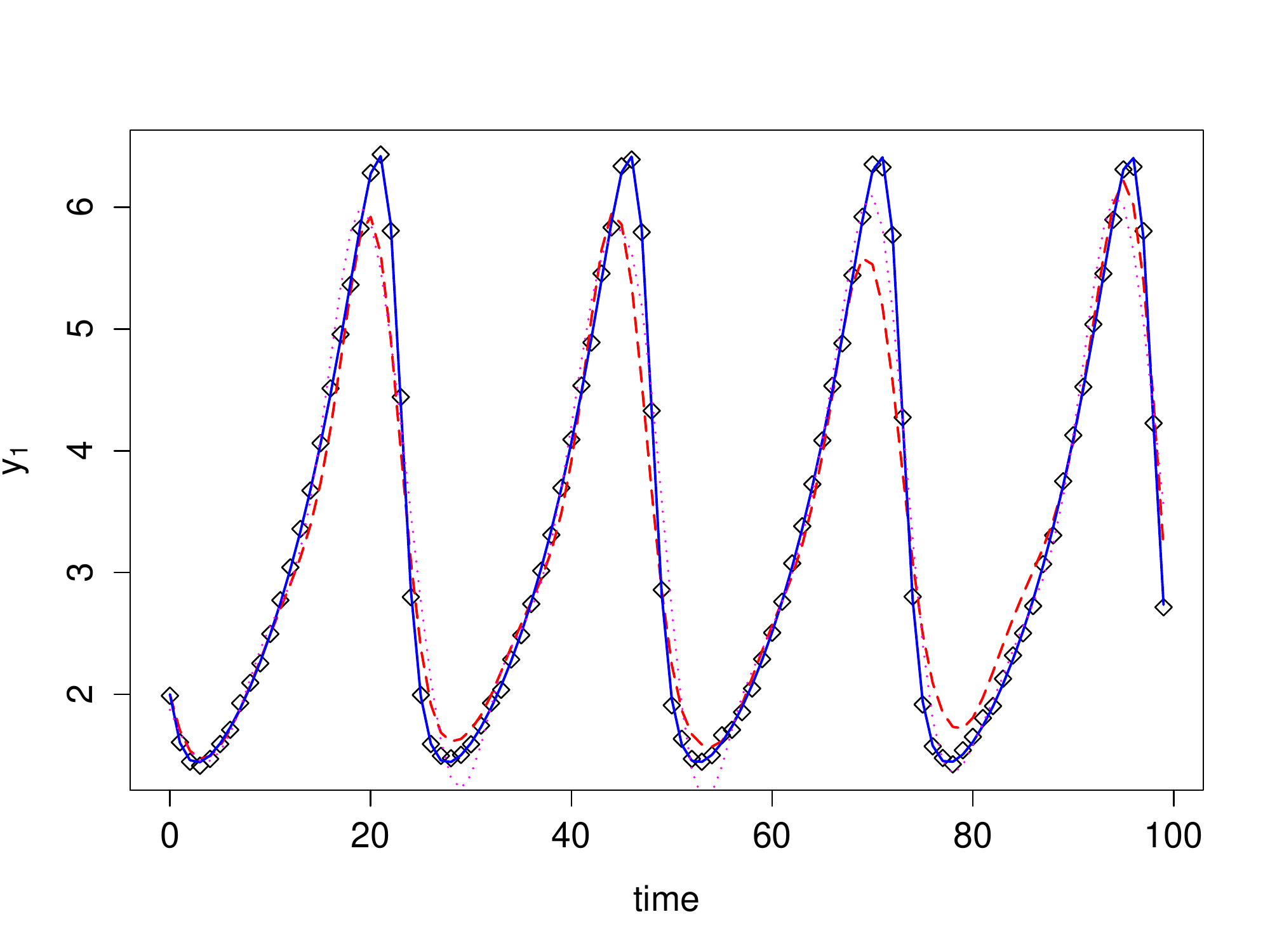} & \includegraphics[scale=0.3]{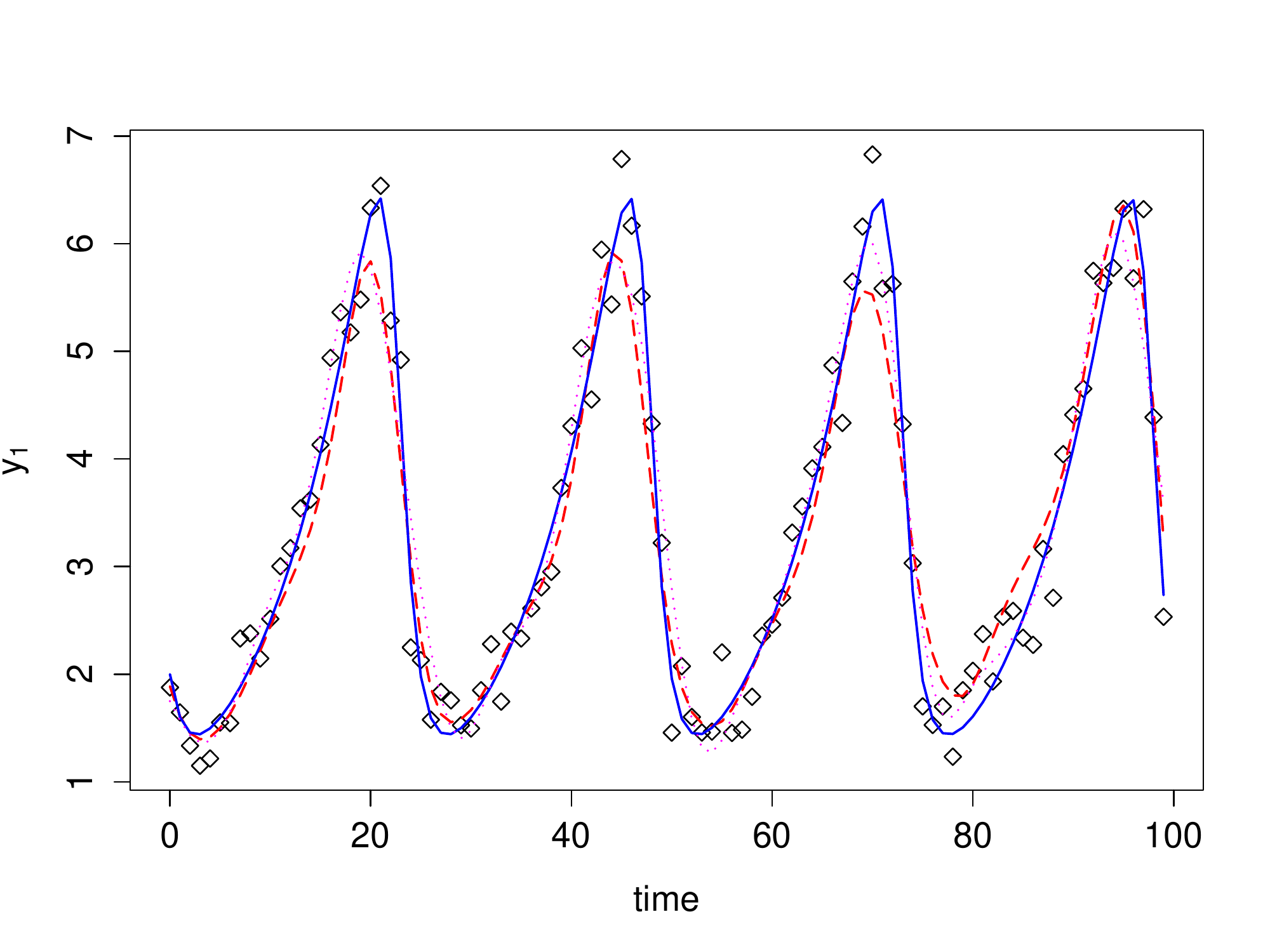} & \includegraphics[scale=0.3]{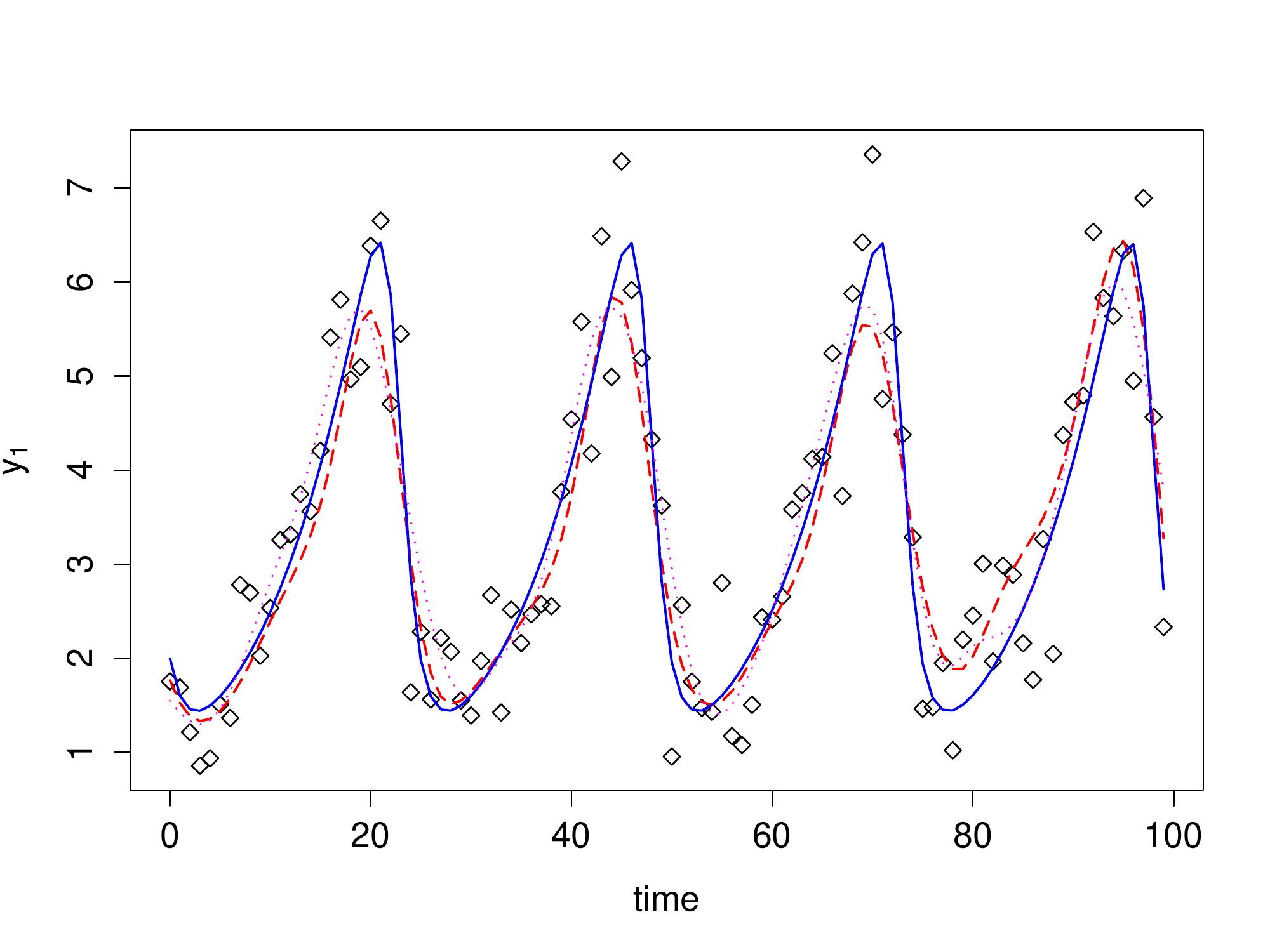} \\ 
\includegraphics[scale=0.3]{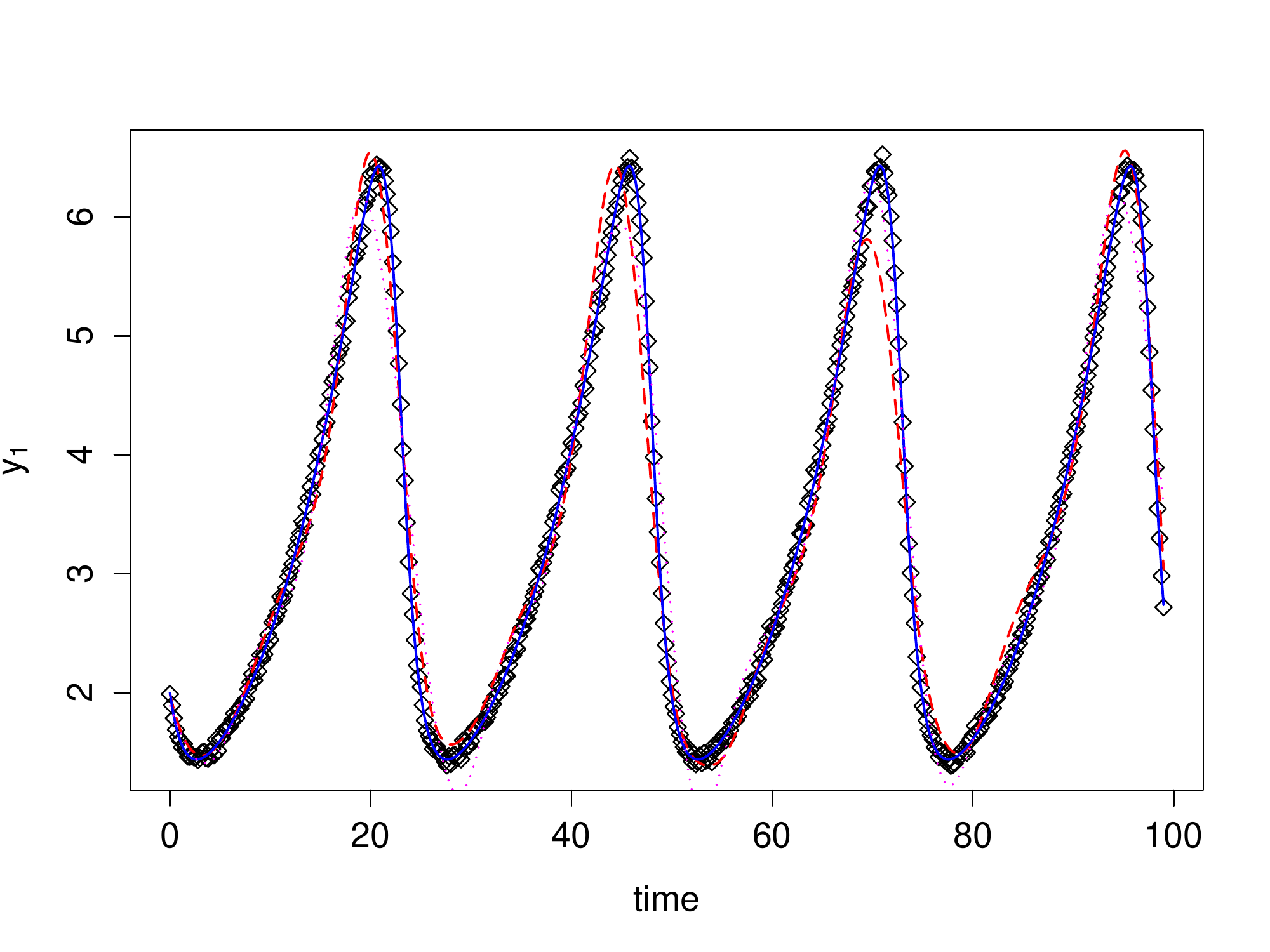} & \includegraphics[scale=0.3]{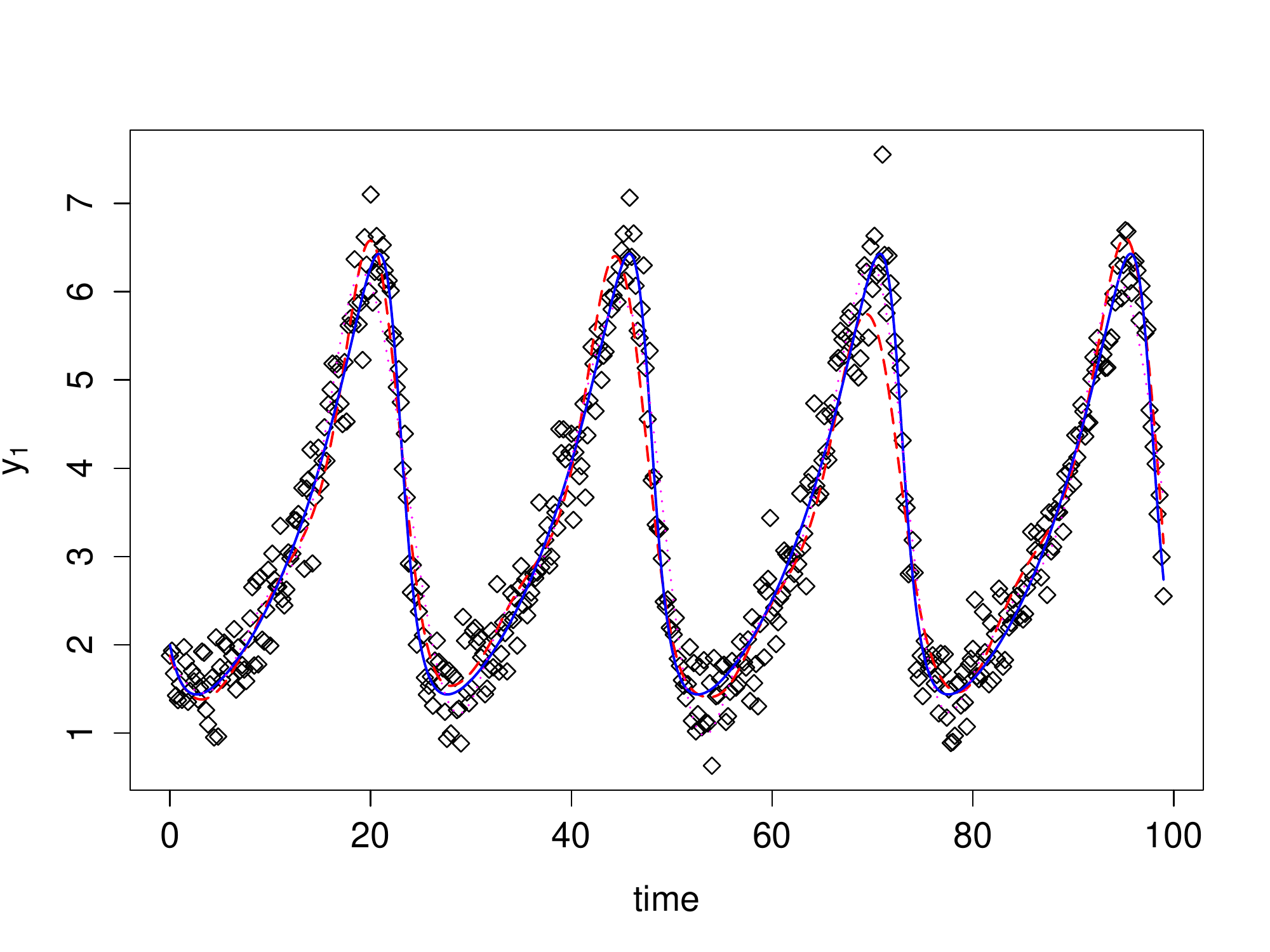} & \includegraphics[scale=0.3]{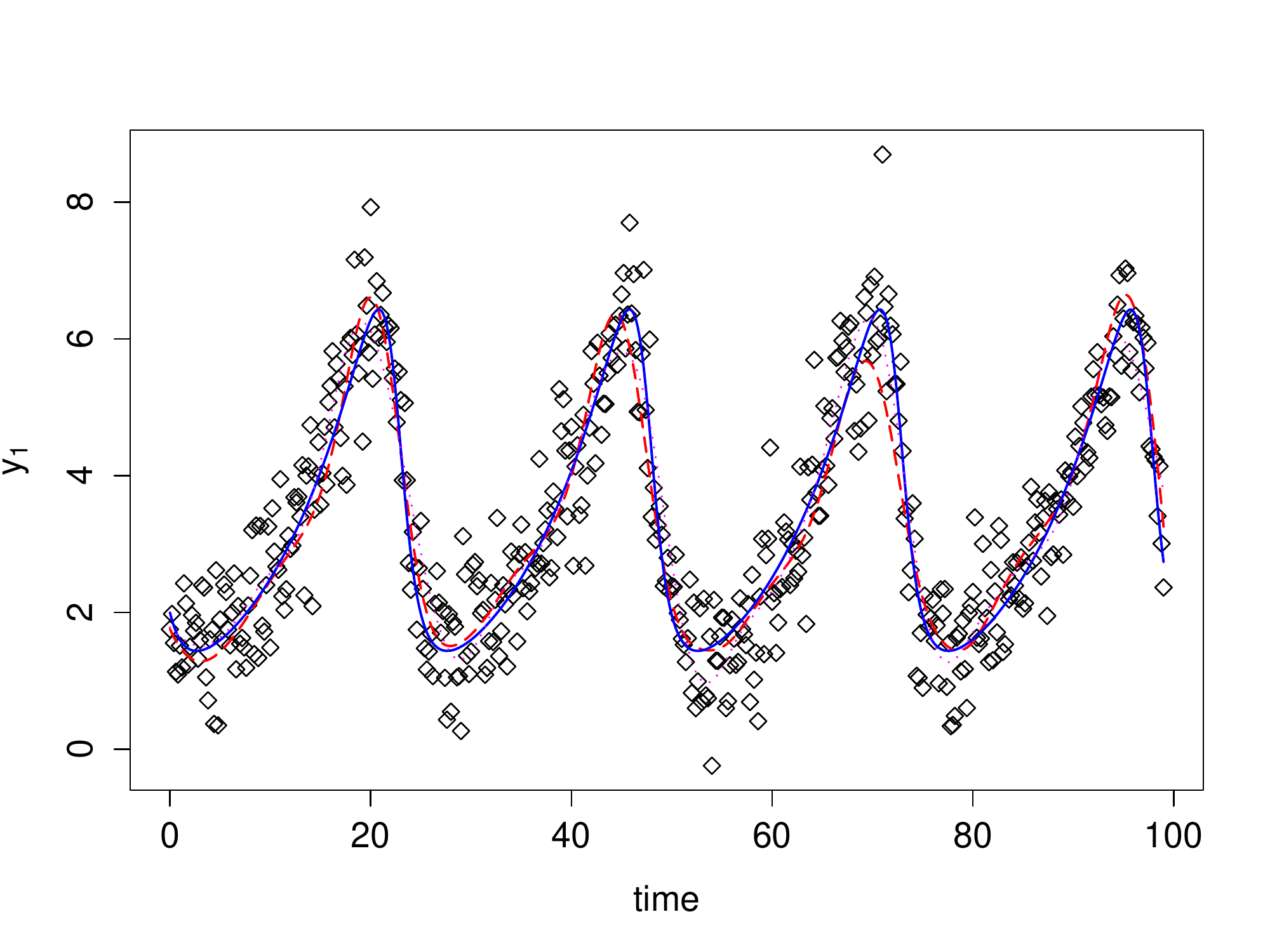} \\ 
\end{tabular}
\caption{Lotka-Volterra simulations: observed noisy data (\emph{dots}), smoothing spline (\emph{dotted} line), true solution of the ODE system (\emph{solid} line) and reconstructed solution (\emph{long-dashed} line) for the first variable $\boldsymbol{x}_{\cdot 1}$. Different sample sizes ($n=25,100,500$) from the top to the bottom and noise levels (low, medium and high) from left to right}\label{lotka_fig}
\end{figure}
\end{landscape}

In the second batch of simulations we consider the Lotka-Volterra system \citep{lotkaref}. This system of ODEs is used to model the dynamics, with respect to time, of two competing groups categorizable as \emph{preys} and \emph{predators}; setting some of the parameters of the ODEs to zero or imposing constraints, however, produces systems that are also used to characterize epidemic processes. The model is described by the following set of equations:
\begin{equation}
\label{lotka_system}
\begin{cases}
\boldsymbol{x}'_{\cdot 1}= \beta_1 \boldsymbol{x}_{\cdot 1} + \beta_2 \boldsymbol{x}_{\cdot 1} \boldsymbol{x}_{\cdot 2} \\
\boldsymbol{x}'_{\cdot 2}=  \beta_3 \boldsymbol{x}_{\cdot 2} + \beta_4 \boldsymbol{x}_{\cdot 1} \boldsymbol{x}_{\cdot 2} \\
x_{11}=\xi_1 \\
x_{12}=\xi_2.
\end{cases}
\end{equation}
We focus the inference procedure on the first ordinary differential equation of the system and thus on the subset of parameters $(\xi_1, \beta_1, \beta_2)$. We explore nine scenarios as in the logistic growth case. We use the same approach to identify the three levels of noise but we select the sample sizes differently: the first one ($n=25$) reproduces a situation where we have 25 evenly-spaced time points starting from $t_1=0$ up to $t_{25}=24$; for $n=100$, $t$ ranges from $t_1=0$ to $t_{100}=99$ with unitary step size; the last one ($n=500$) has the same time range as $n=100$, with $t_1=0$ and $t_{500}=99$, but a denser sampling grid given by 0.2 as the step size. Going from $n=100$ to $n=500$ encompass a situation where the time range is fixed (the maximum observational time point) but the amount of data increases (more observations in the same timeframe). We use as regressing functions the quantities $h_1(\boldsymbol{x})=\boldsymbol{x}_{\cdot 1}$ and $h_2(\boldsymbol{x})=\boldsymbol{x}_{\cdot 1} \boldsymbol{x}_{\cdot 2}$; we choose the same number of knots, $q_1=q_2=5$, for both splines. In Table \ref{lotka_tab}, averages of the posterior means and corresponding mean square errors are reported for all the scenarios. We see that both algorithms perform well when the sample size is $n=25$, regardless of the three noise level (SNR=50,5,2.5). The number of actual datasets used to compute averages for \emph{CollocInfer} also shows that convergence was achieved for all the first three scenarios. When evaluating the performance of the two methods, for $n=100$, we notice that \emph{SnM} performs slightly better in terms of bias of the estimated parameters $\beta_1$ and $\beta_2$; also, \emph{CollocInfer} shows slightly higher MSEs for the second estimated parameter with respect to \emph{SnM}. This difference is more prominent when $n=500$. As expected, the quantification of the uncertainty about the solution provided by \emph{SnM}, through $\sigma^2_1$, decreases as the sample size grows. A visual appreciation of the results is given in Figure \ref{lotka_fig}. The most different performance between the smoothing and the regression curves is evident in the case of $n=25$ and $SNR=2.5$ (upper row of the mosaic, rightmost plot): in this case, the mean square errors of the dotted line and the long-dashed line are respectively $\mbox{MSE}_{x_1}=0.690$ and $\mbox{MSE}_{g_1}=0.252$. Such a discrepancy in the (average) accuracy of the reconstructed curves fades as the sample size grows. For $n=500$ and $SNR=50$ (the lowest noise level), the two errors are $\mbox{MSE}_{x_1}=0.094$ and $\mbox{MSE}_{g_1}=0.084$.

\subsection{HIV viral fitness}

\begin{landscape}
\begin{figure}
\begin{tabular}{lll}
\includegraphics[scale=0.3]{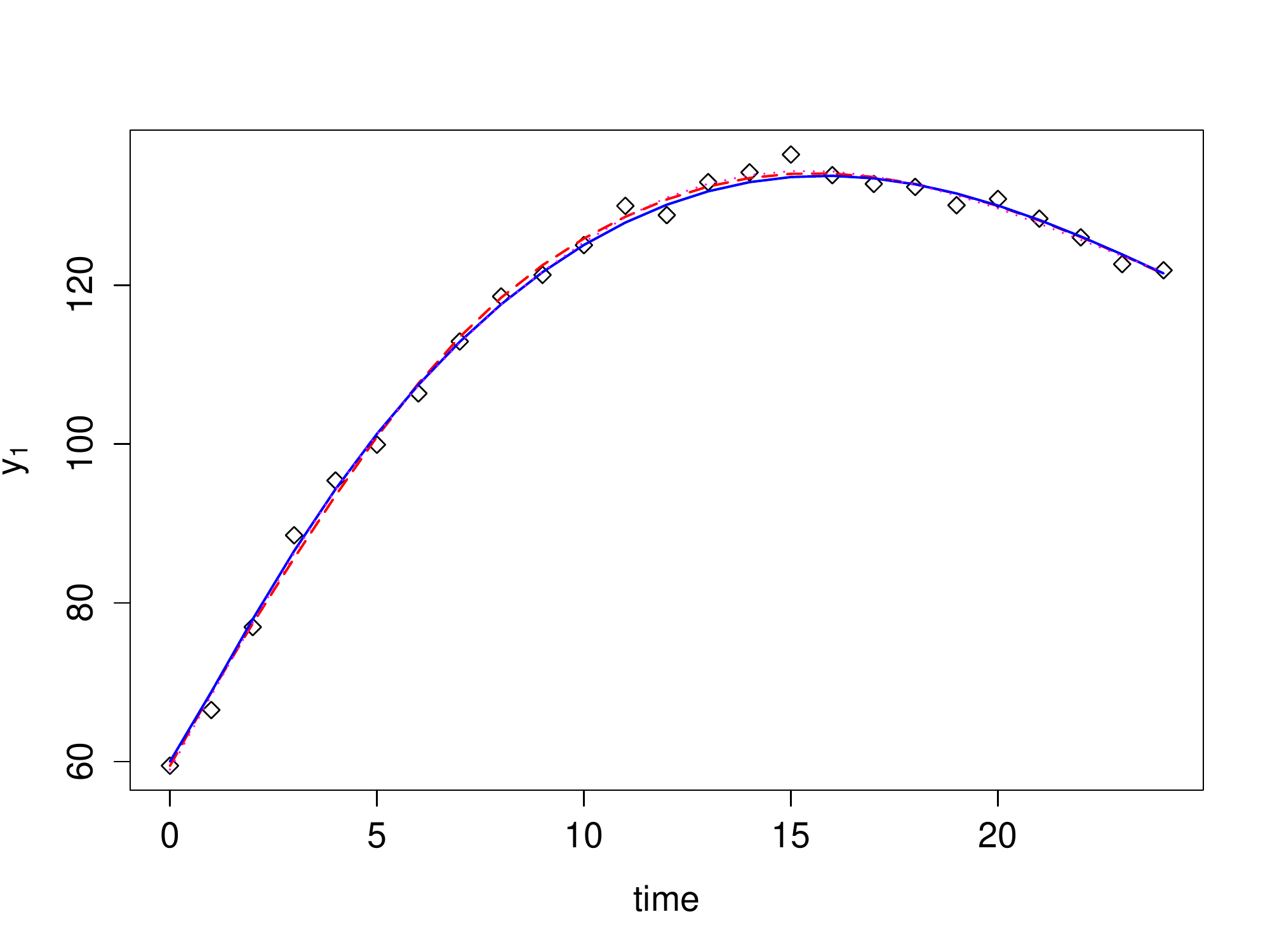} & \includegraphics[scale=0.3]{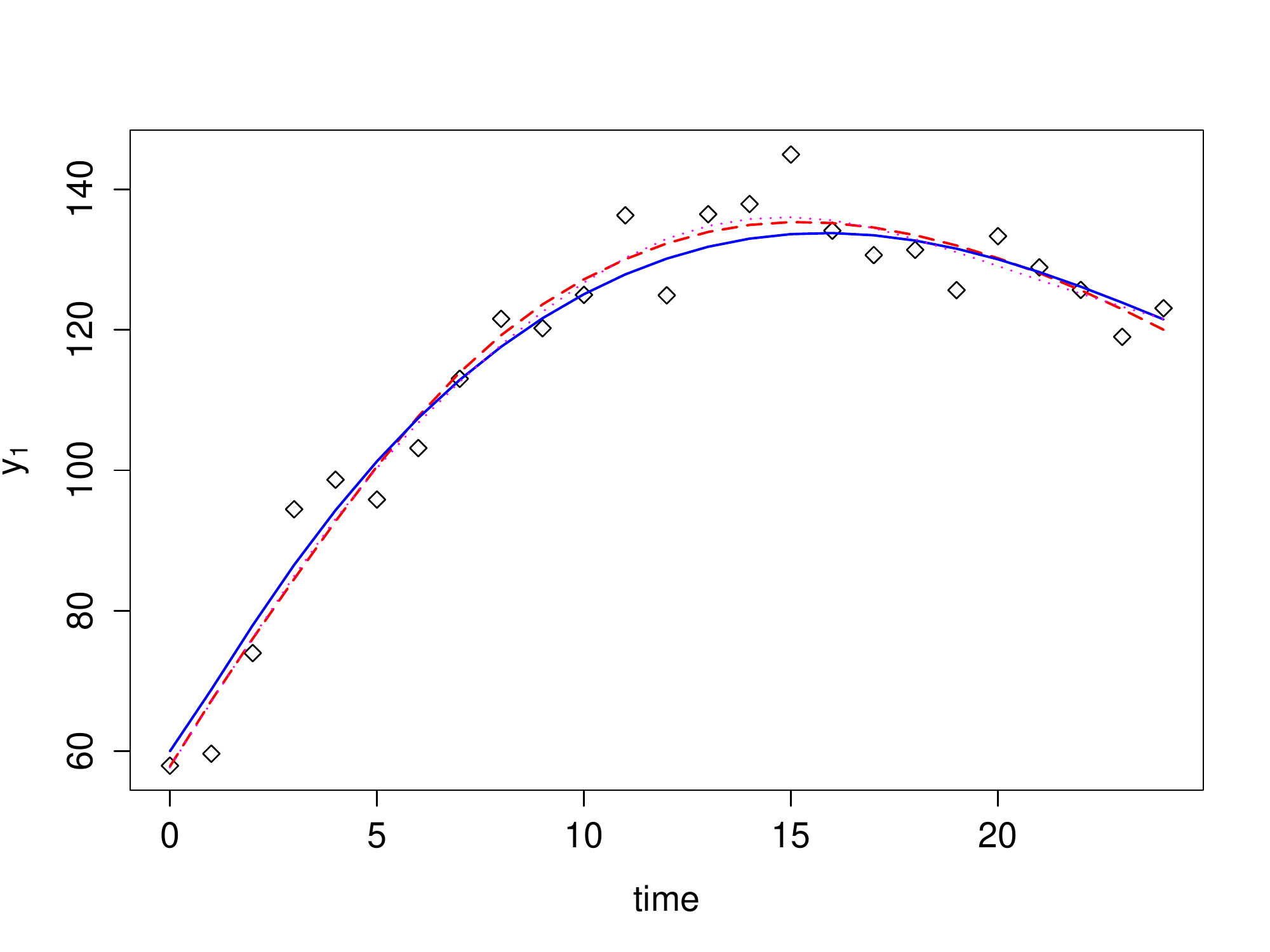} & \includegraphics[scale=0.3]{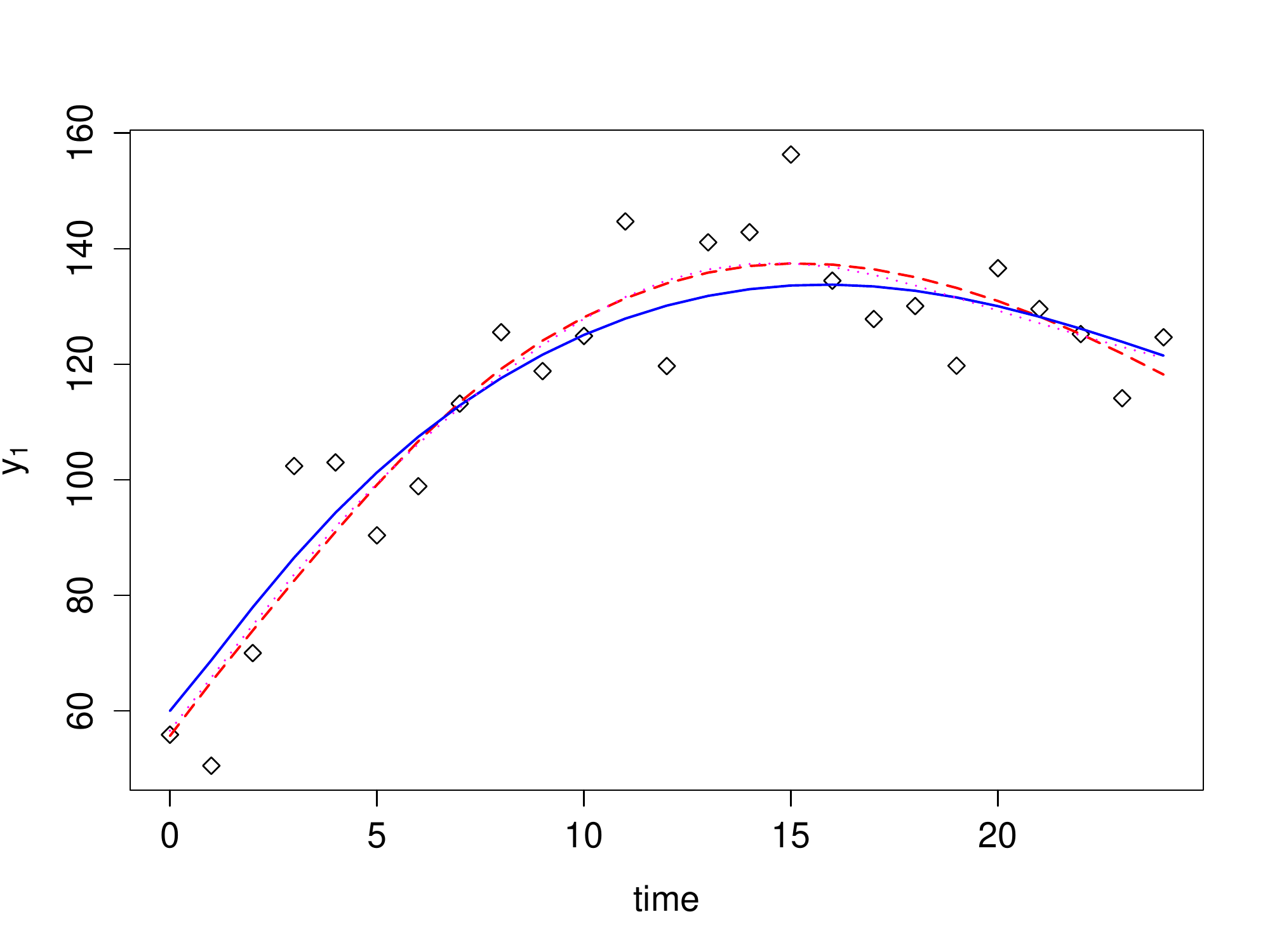} \\ 
\includegraphics[scale=0.3]{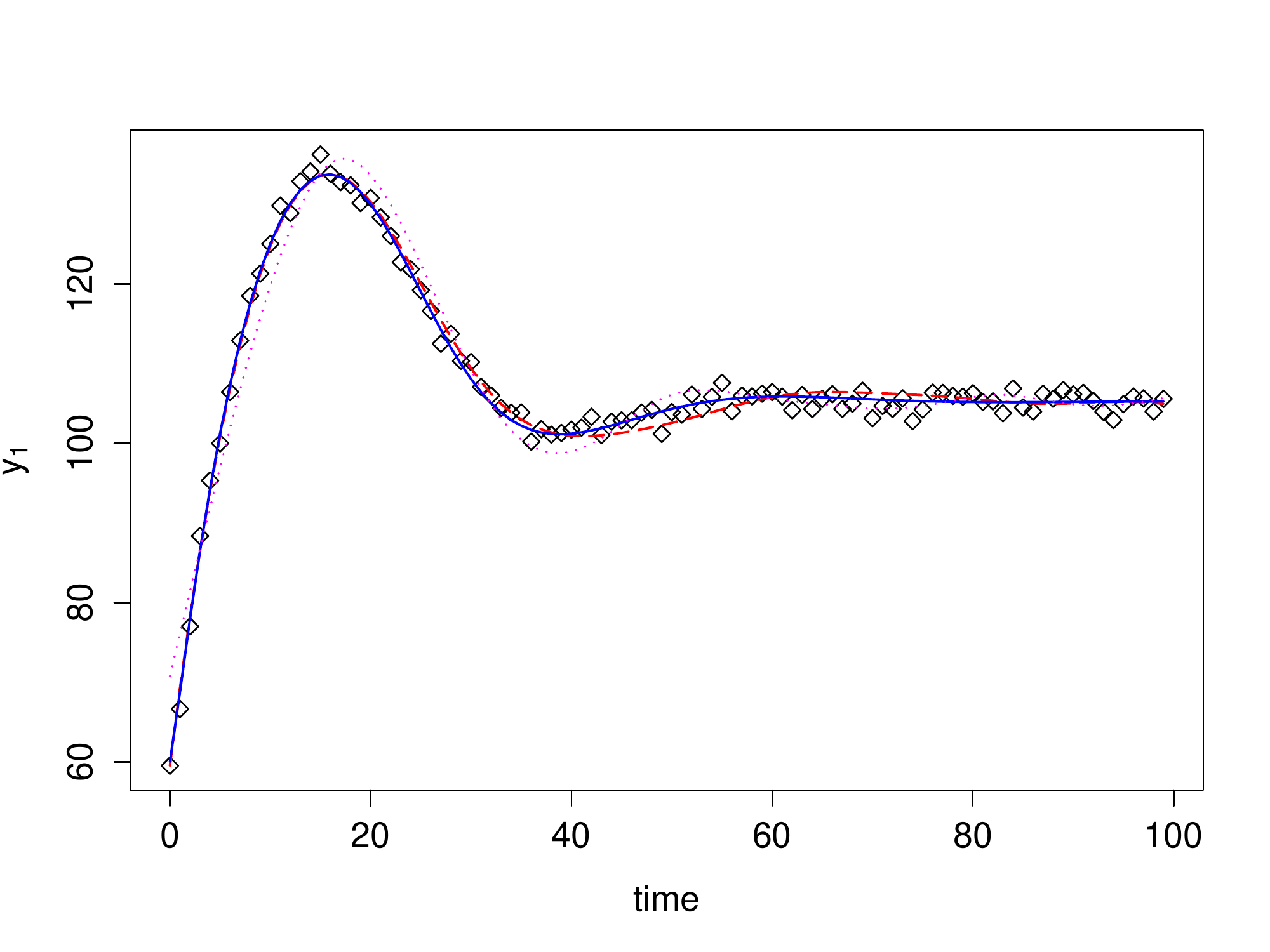} & \includegraphics[scale=0.3]{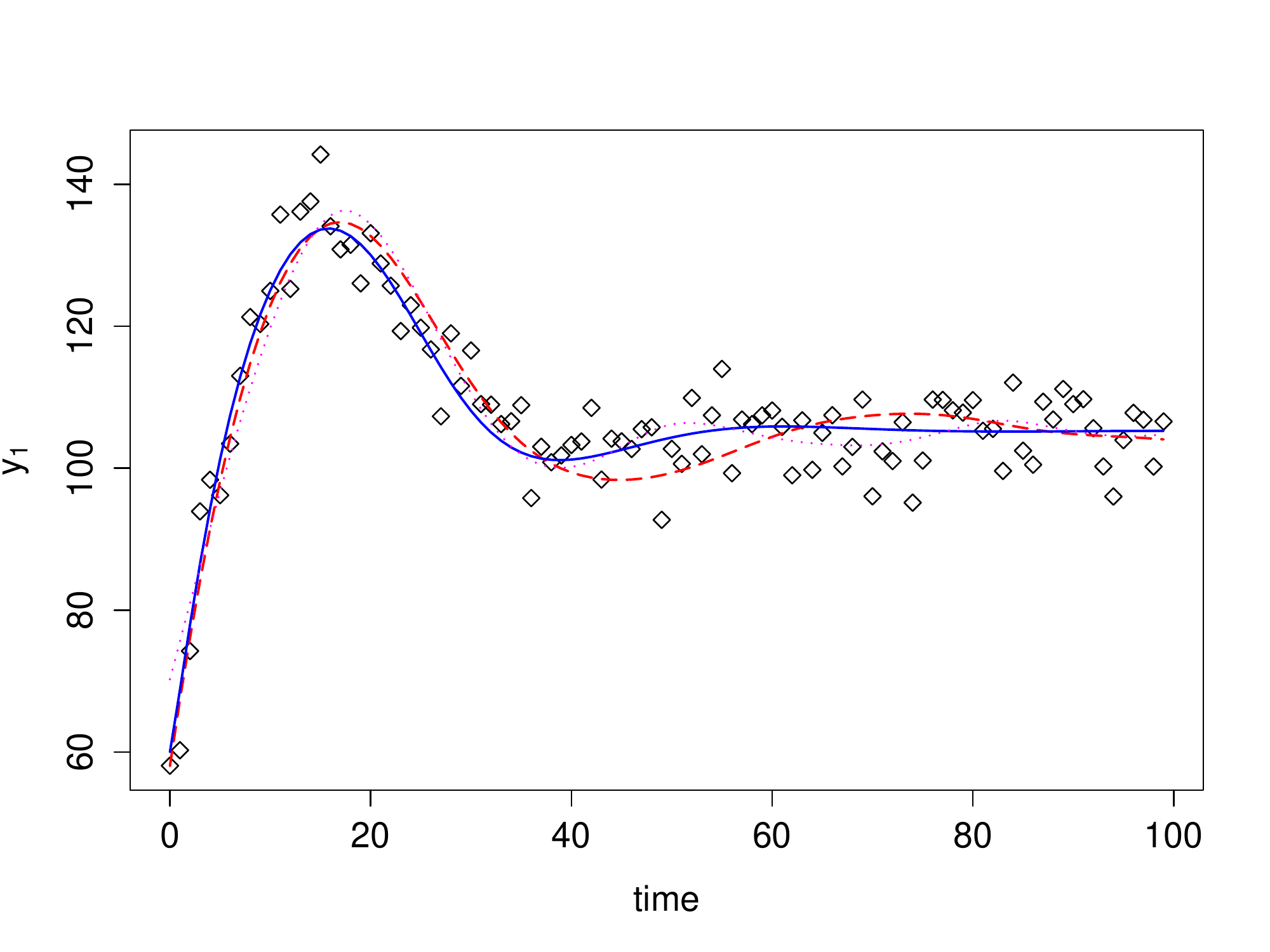} & \includegraphics[scale=0.3]{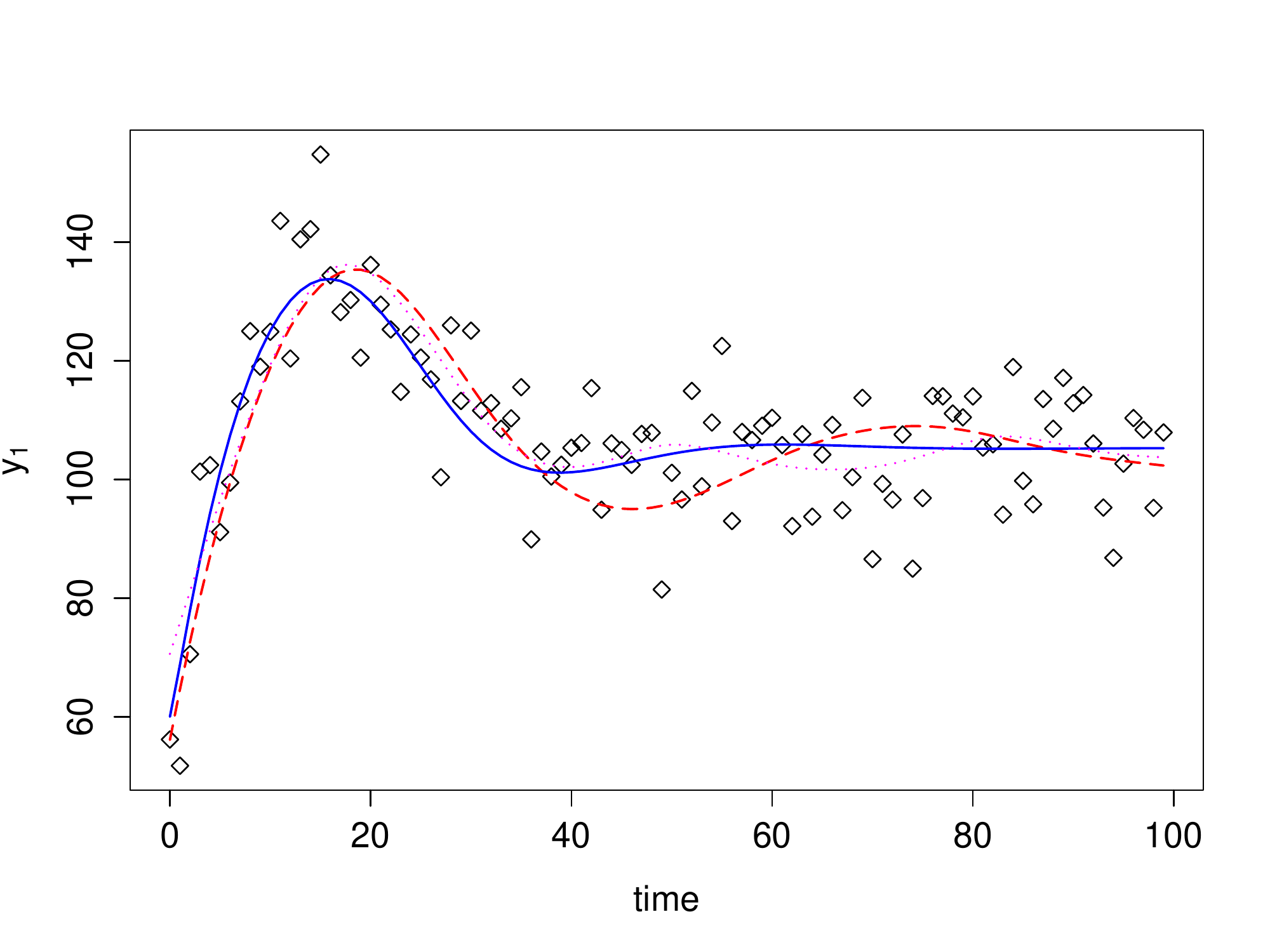} \\ 
\includegraphics[scale=0.3]{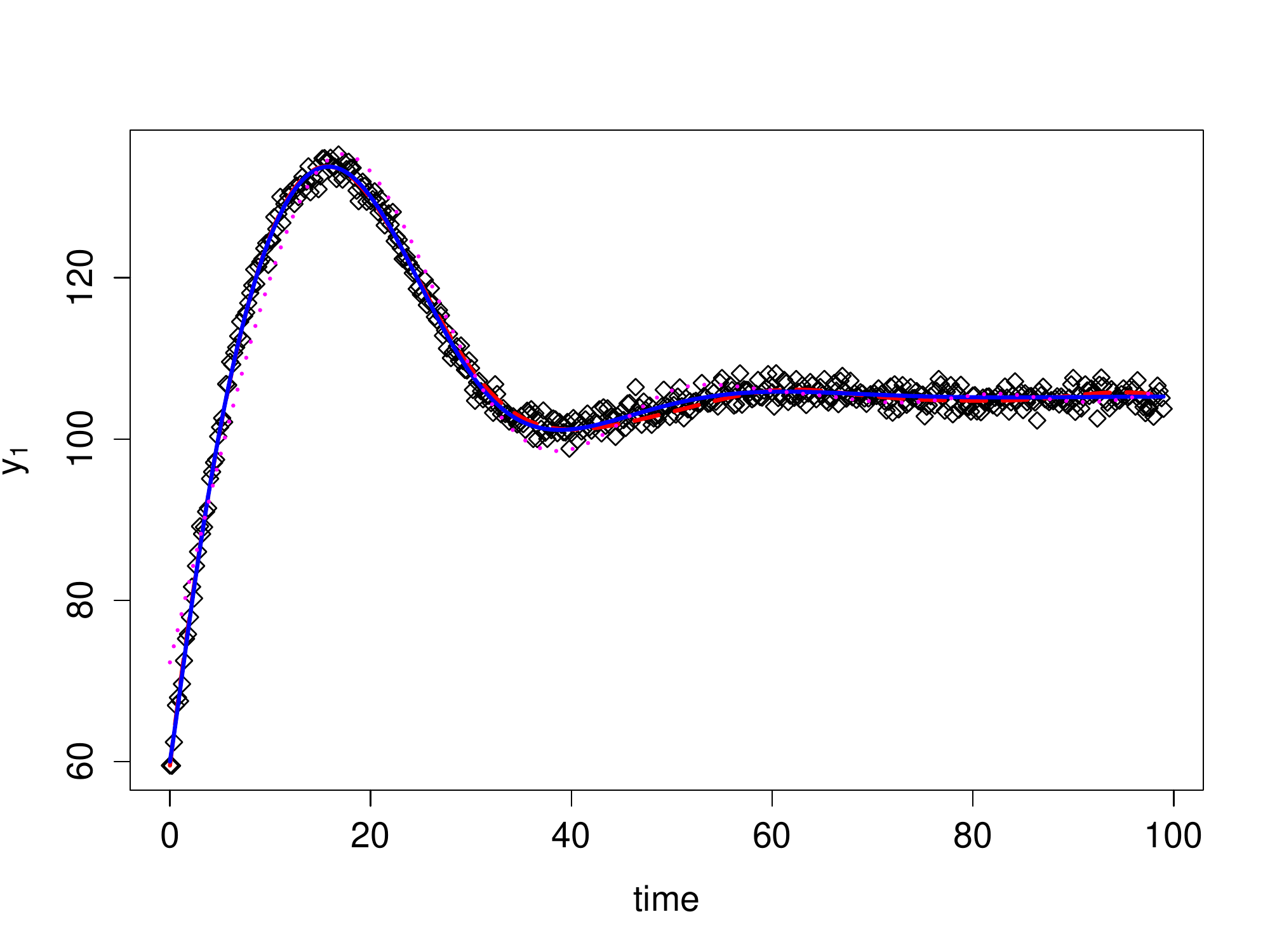} & \includegraphics[scale=0.3]{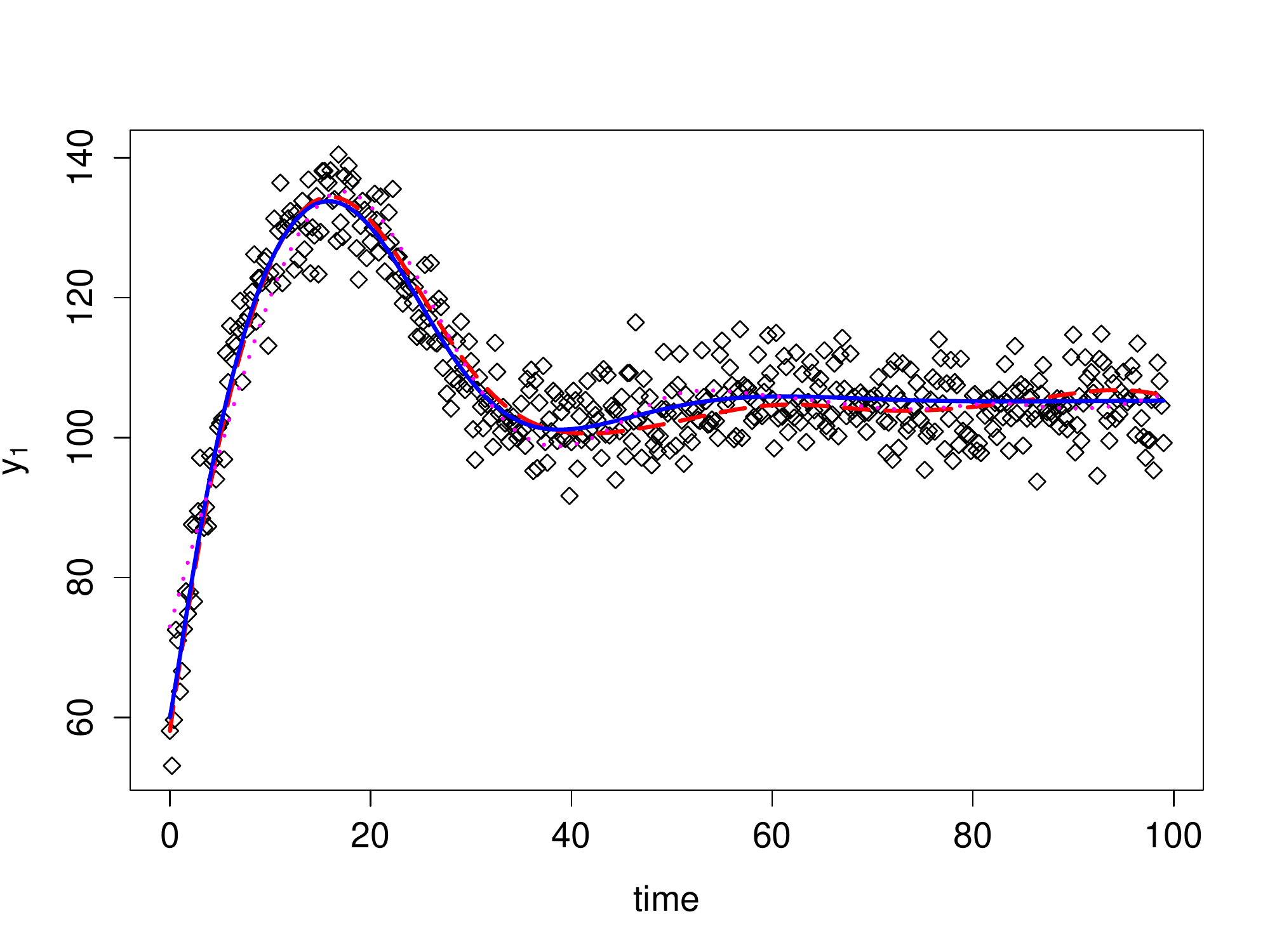} & \includegraphics[scale=0.3]{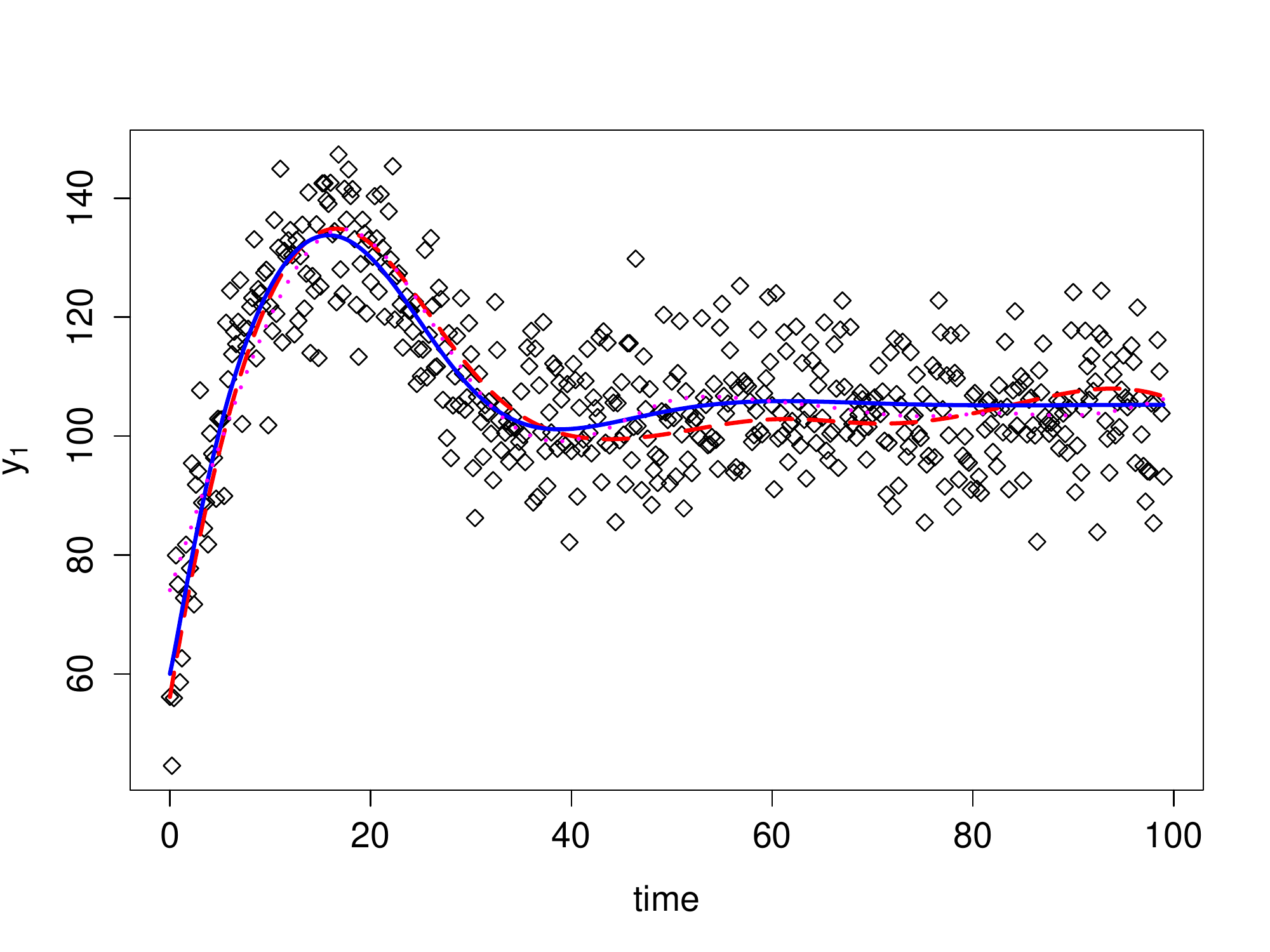} \\ 
\end{tabular}
\caption{HIV viral fitness simulations: observed noisy data (\emph{dots}), smoothing spline (\emph{dotted} line), true solution of the ODE system (\emph{solid} line) and reconstructed solution (\emph{long-dashed} line) for the first variable $\boldsymbol{x}_{\cdot 1}$. Different sample sizes ($n=25,100,500$) from the top to the bottom and noise levels (low, medium and high) from left to right}\label{hiv_fig}
\end{figure}
\end{landscape}

Our last batch of data is simulated from a set of ODEs modelling the dynamics of HIV virus \citep{bonhoeffer1997virus}. The system is defined as:
\begin{equation}
\label{hiv_system}
\begin{cases}
\boldsymbol{x}'_{\cdot 1}= \beta_1 + \beta_2 \boldsymbol{x}_{\cdot 1} + \beta_3 \boldsymbol{x}_{\cdot 1} \boldsymbol{x}_{\cdot 3} \\
\boldsymbol{x}'_{\cdot 2}= \beta_3 \boldsymbol{x}_{\cdot 1} \boldsymbol{x}_{\cdot 3} - 0.5 \boldsymbol{x}_{\cdot 2} \\
\boldsymbol{x}'_{\cdot 3}= 0.5 \cdot \beta_4 \boldsymbol{x}_{\cdot 2}  + \beta_5 \boldsymbol{x}_{\cdot 3}\\
x_{11}=\xi_1\\
x_{12}=\xi_2\\
x_{13}=\xi_3.
\end{cases}
\end{equation}
where we focus on the first ODE and on the subset of parameters $(\xi_1, \beta_1, \beta_2, \beta_3)$. The simulation setting is the same as in Section \ref{lotka_sec}: three levels of noise and three sample sizes where only the last has an effective increase in the number of observations. True parameters' values are tuned according to the ones used in \citet{Vujacic}. Number of knots selected is the same for all splines $q_1=q_2=q_3=20$. We use the following regressing functions: $h_1(\boldsymbol{x})=1$, $h_2(\boldsymbol{x})=\boldsymbol{x}_{\cdot 1}$ and $h_3(\boldsymbol{x})=\boldsymbol{x}_{\cdot 1} \boldsymbol{x}_{\cdot 3}$. We report the results in Table \ref{hiv_tab} and Figure \ref{hiv_fig}. The parameter $\beta_2$ proved to be difficult to estimate with \emph{CollocInfer} so we decided to fix it to the true value for this algorithm while estimating it in the case of \emph{SnM}. When $n=25$ we notice that our method breaks down at the highest level of noise contamination (SNR=1.3): set aside the recovered initial condition, which shows good average posterior mean, the estimates for the other parameters are noticeably biased and have large mean square errors (especially in the case of $\beta_1$). Even if the NAD for \emph{CollocInfer} is around 85 for the first three scenarios, the algorithm seems to be more robust in this setting. When considering $\beta_1$ and $\beta_3$, \emph{CollocInfer} achieves lower mean square errors in almost every scenario with the exception of the setting with $n=500$ and SNR=15, where the MSE for $\beta_1$ obtained with \emph{SnM} is slightly lower than the one computed for \emph{CollocInfer}. The number of discarded datasets gets lower as the sample size grows, as expected. For $n=100$ and $n=500$ we do not observe the same degrading effect on our algorithm \emph{SnM} even at the highest level of noise, meaning that the information from the bigger sample size is enough to overcome the contamination in the data. As we can see from the mosaic plot in Figure \ref{hiv_fig}, for $n=100$ and $n=500$ there is an appreciable difference between the two curves with respect to the true one, especially on the right-half portion of each plot. The average mean square errors for the setting with $n=100$ and SNR=1.3 are $\mbox{MSE}_{g_1}=16.179$ and $\mbox{MSE}_{x_1}=11.743$, meaning that the smoothing step provides a reconstructed curve slightly closer to the true one (dotted line versus solid line). The opposite happens when the sample size gets to 500, as in the setting with SNR=2.5, the two MSEs are and $\mbox{MSE}_{g_1}=1.397$ and $\mbox{MSE}_{x_1}=5.859$. As for the setting in which the model breaks down ($n=25$, SNR=1.3), the average mean square errors are $\mbox{MSE}_{g_1}=80.467$ and $\mbox{MSE}_{x_1}=104.969$, values much higher than the ones computed on the other eight scenarios (not reported here for brevity).

\subsection{FitzHugh-Nagumo system for neuron electrical activity} \label{nagumo_sec}

We analyze a toy-data example available from the package \emph{CollocInfer} \citep{hooker2010collocinfer}. The data (\emph{FhNdata}) consist of 41 evenly-spaced observations in the timeframe $[0,20]$ from the following ODEs model 
\begin{equation}
\label{nagumo_eq}
\begin{cases}
\boldsymbol{x}'_{\cdot 1}= c(\boldsymbol{x}_{\cdot 1}-\boldsymbol{x}^3_{\cdot 1}/3 + \boldsymbol{x}_{\cdot 2})\\
\boldsymbol{x}'_{\cdot 2}= -\frac{1}{c}(\boldsymbol{x}_{\cdot 1}-a+b\boldsymbol{x}_{\cdot 2}) \\
x_{11}=\xi_1 \\
x_{12}=\xi_2
\end{cases}
\end{equation}
known as the FitzHugh-Nagumo system \citep{fitzhugh1961impulses,nagumo1962active}. The set of equations describe the pulse transmission for neuronal activity. The parameters' values used to generate the data are $a=0.2, d=0.2, c=3, \xi_1=0.5$; the simulated values are then perturbed with variances equal to 0.25 for both the variables $\boldsymbol{x}_{\cdot 1}$ and $\boldsymbol{x}_{\cdot 2}$. As it is written in Equation \ref{nagumo_eq}, the system is not linear in the parameters. We thus consider another representation
\begin{equation}
\begin{cases}
\boldsymbol{x}'_{\cdot 1}= \beta_4(\boldsymbol{x}_{\cdot 1}-\boldsymbol{x}_{\cdot 1}^3/3+\boldsymbol{x}_{\cdot 2}) \\
\boldsymbol{x}'_{\cdot 2}= \beta_1 \boldsymbol{x}_{\cdot 1} + \beta_2 + \beta_3 \boldsymbol{x}_{\cdot 2}\\
x_{11}=\xi_1 \\
x_{12}=\xi_2
\end{cases}
\end{equation}
and we focus on the second differential equation and the subset of parameters $(\xi_2, \beta_1=-1/c,\beta_2=a/c,\beta_3=-d/c)$. The ODE solution regression uses the following functions: $h_1(\boldsymbol{x})=\boldsymbol{x}_{\cdot 1}$, $h_2(\boldsymbol{x})=1$ and $h_3(\boldsymbol{x})=\boldsymbol{x}_{\cdot 2}$. We compare our results with the point estimates obtained by \citet{Vujacic}, remarking the fact that we only do inference on one of the two differential equations whereas they simultaneously estimate all the parameters of the model using information from both variables. In Figure \ref{nagumo_fig}, reconstructed curves from both steps of the procedure (\emph{smooth} and \emph{match}) are plotted:  we can see, in the right plot, the penalized spline (\emph{dotted} line) captures a biased initial condition in comparison to the ODE solution regression (\emph{long-dashed} line), even if the associated mean square errors for the two curves, respectively $\mbox{MSE}_{g_1}=0.042$ and $\mbox{MSE}_{x_1}=0.033$, are actually close. The degree of smoothing for the two curves is also practically the same. The smoothing for the first variable (left plot of Figure \ref{nagumo_fig}) is satisfactory. As for the other parameters, we report their posterior means in Table \ref{nagumo_tab}: the algorithm recovers the true value for $\beta_1$ with appreciable accuracy; for $\beta_2$ and especially $\beta_3$, the algorithm returns slightly biased posterior means. Although, as stated before, there is no theoretical true correspondence between the estimated shared nuisance parameter and the variance of the noise added to the data, the posterior mean for $\sigma^2_2$ is lower than the one used to perturb the data, meaning that - potentially - the additional information from the ODE solution helps shrinking down the overall measure of uncertainty regarding the second variable $\boldsymbol{x}_{\cdot 2}$ that we are trying to model.

\begin{table*}[ht!]
\centering
\caption{Results for the \emph{FhNdata}: posterior means (\emph{posterior standard deviations} within brackets) from \emph{SnM} in the second column and point estimates from \citet{Vujacic} in the third column}\label{nagumo_tab}
\begin{tabular}{lcc}
\\
\hline\hline
Parameters & Post. Mean (\emph{post. sd}) & \citet{Vujacic} \\
\hline
$\xi_2=0.5$ & 0.696 (\emph{0.244}) & 0.569 \\
$\beta_1=-0.33$ & -0.322 (\emph{0.041}) & -0.333 \\
$\beta_2=0.067$ & 0.091 (\emph{0.025}) & 0.106 \\
$\beta_3=-0.067$ & -0.028 (\emph{0.066}) & -0.047 \\
\hline
$\sigma^2_2$ & 0.060 (\emph{0.019})& \\
\hline
\multicolumn{2}{l}{\begin{footnotesize}Runtime: 10,000 MCMC iterations in 29.17 seconds.\end{footnotesize}}
\end{tabular}
\end{table*}

\begin{figure*}[ht!]
\centering
\begin{tabular}{ll}
\includegraphics[scale=0.5]{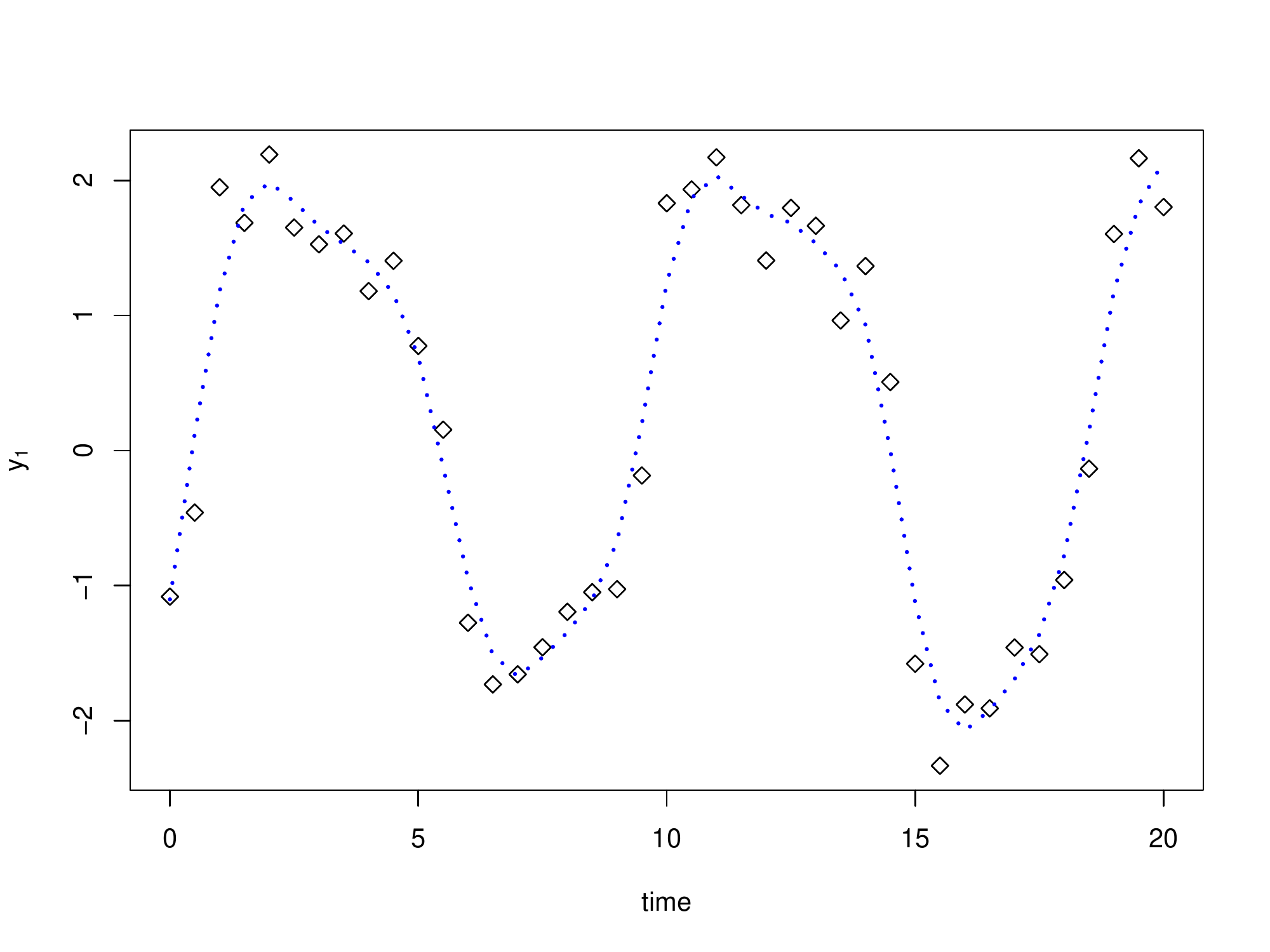} \\ \includegraphics[scale=0.5]{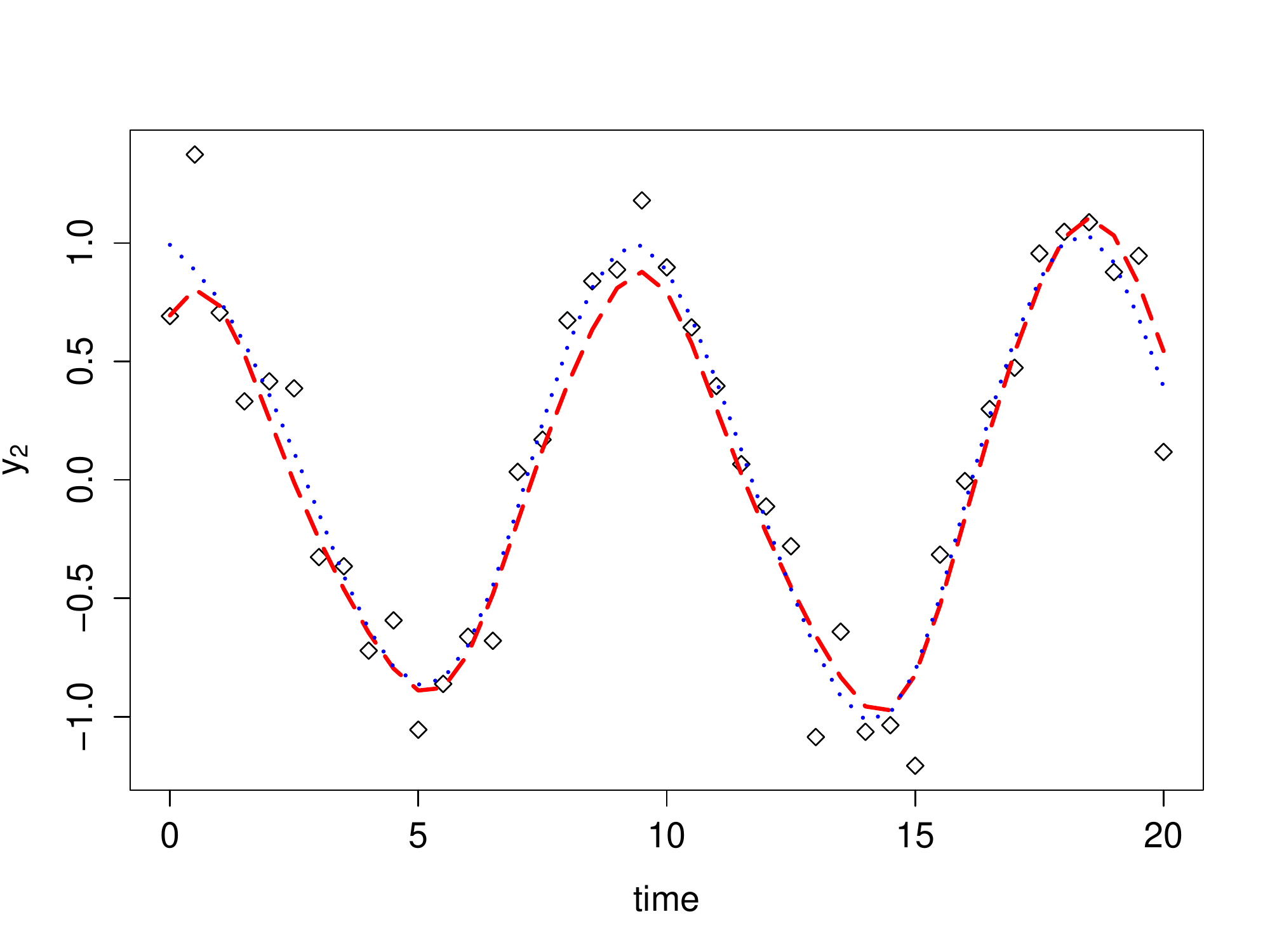}
\end{tabular}
\label{nagumo_fig} 
\caption{FitzHugh-Nagumo system for neuron electrical activity: observed noisy data (\emph{dots}), smoothing spline (\emph{dotted} line) and reconstructed solution (\emph{long-dashed} line, only right plot) for the first variable (left plot) and the second variable $\boldsymbol{x}_{\cdot 2}$ (right plot)} \label{nagumo_fig}
\end{figure*}

\newpage

\section{Conclusions}\label{conc_sec}

We have proposed a Bayesian approach to indirectly solve an equation of an ODEs' system while doing inference on its parameters. The employed strategy is suitable for models that are linear in the parameters, with observations available for all the components of the system. It is compartmentalized into two main stages: a first \emph{smoothing} step, that serves as a reconstruction of the components of the process through penalized spline smoothing of the noisy observed data; a second \emph{match} step, where the smoothed curves are numerically integrated and used as inputs for ridge regression. The two phases of the procedure are jointly governed by  $\sigma^2_k$, a noise parameter - common to both steps - that measures the solution uncertainty of the $k$-th equation of the system we are indirectly solving. This parameter brings together two main sources of uncertainty: measurement error and model error. We evaluated the performance and reliability of the strategy through different ODEs systems, starting from a simple one (with only one variable) and then moving to processes that had two or three variables and more complex time dynamics. We also tested the approach on a dataset previously analyzed by \citep{Vujacic}, to compare the results. The procedure we propose has the advantages of being fast, simple to implement, and provides the ODE solution as a by-product of the inference procedure. The `tuning' parameters are minimal: the number of knots and their placement have no substantial impact on the reliability of the smoothing step; different splines (i.e, \emph{B-spline}, \emph{thin plates}, etc.), that do not actually require such a choice, can be employed in the first step to address the issue. An interesting alternative would be to employ the P-splines approach proposed in \citet{ventrucci2016}. In their work, the authors use a penalized complexity prior, that is a (prior) distribution injecting information not in terms of the penalizing term $\lambda_{\theta_k}$ but as our prior guess about the polynomial order needed to reconstruct $\boldsymbol{x}_k$ and how likely would be to adopt a higher order polynomial. This is quite an appealing approach because it is usually easier to elicit the information in terms of equivalent polynomial order, especially in ODEs systems context. As far as the integration is concerned, we rely on an easy to implement - albeit `rough' - trapezoidal rule that uses the observed time points as the grid to evaluate the integral: a better approximation can be achieved by employing a finer grid, at the cost of increased computational times. Other types of penalization, instead of the ridge, could be explored for the regression step of the strategy, losing however the correspondence with the Tikhonov regularization. 

For future developments, we aim to be able to estimate all the parameters of the system while indirectly solving together and simultaneously the ordinary differential equations, instead of focusing on one of them. As trivial as it may seems to extend the approach, careful consideration is needed before moving toward this direction. For example, a first idea would be to independently run the procedure for each equation but, in that case, we would not be truly using the information at our disposal about the relationships between the variables. The disjoint approach to the set of ODEs could potentially be valuable in a parallelization framework where there are a huge number of equations and a fast explorative result is needed. Another issue would be, when considering all the equations together, which of the two reconstructed curves (the smoothed spline and the regression solution) to use at each iteration of the MCMC procedure given that the measure of uncertainty considers them both. An interesting extension of the model could consider the regressing functions $h_j(\boldsymbol{x})$ as not known in advance and to be estimated along with the other parameters of the model. We already briefly explored this aspect using a Bayesian smoothed spline regression at the \emph{match} step of the procedure: we obtained a satisfactory reconstruction of the curve but at the expense of losing interpretability of the parameter vector $\boldsymbol{\beta}$, meaning that further investigation on this topic is needed. Finally, as seen in Section \ref{mimic}, an interesting aspect would be to investigate deeper (and potentially quantify, theoretically) the consequences of the shared nuisance parameter $\sigma^2_1$. In this direction, an option could be to express the two levels of noise from the two sources of information ($\boldsymbol{y}_{\cdot 1}$ and $\boldsymbol{y}_{1}^{\star}$) as a proportion of the total shared nuisance parameter and to decide which curve to use, the smoothed one or the system's solution, based on these fractions.

%

\newpage



\section*{References}
  \bibliographystyle{elsarticle-harv} 
  \bibliography{ode_biblio}

\newpage

\begin{landscape}
\begin{table*}[p]
\centering
\caption{Average posterior means (with \emph{MSE} in brackets) for the parameters of the logistic population growth model; three sample sizes and increasing noise level}\label{growth_tab}
\begin{tabular}{cl|cc|cc|cc}
\\
\hline\hline
  & \multirow{3}{*}{Parameters} & \multicolumn{6}{c}{Noise level} \\
  \cline{3-8}
 Sample & & \multicolumn{2}{c|}{low ($SNR=13$)} & \multicolumn{2}{c}{medium ($SNR=6.5$)} & \multicolumn{2}{|c}{high ($SNR=1.3$)} \\
 size & & SnM & CollocInfer & SnM & CollocInfer & SnM & CollocInfer \\ 
\hline
\multirow{3}{*}{$n=25$} & $\xi_1=0.1$ & 0.098 (\emph{0.001}) & - & 0.096 (\emph{0.002}) & - & 0.078 (\emph{0.043}) & - \\
& $\beta_1=2.5$ & 2.531 (\emph{0.105}) & 3.819 (\emph{6.845}) & 2.560 (\emph{0.412}) & 5.054 (\emph{14.530}) & 2.317 (\emph{4.687}) & 4.727 (\emph{12.370}) \\
& $\beta_2=-0.125$ & -0.180 (\emph{0.271}) & -0.244 (\emph{0.091}) & -0.216 (\emph{1.035}) & -0.420 (\emph{0.228}) & -0.359 (\emph{9.864}) & -0.385 (\emph{0.215}) \\[1ex]
&$^\diamond \sigma^2_1$ & 0.064  & - & 0.244  & - & 7.054  & - \\
& NAD & 100 & 98 & 100 & 96 & 100 & 93 \\
\hline
\multirow{3}{*}{$n=100$} & $\xi_1=0.1$ & 0.098 (\emph{0.001}) & - & 0.096 (\emph{0.002}) & - & 0.078 (\emph{0.046}) & - \\
& $\beta_1=2.5$ & 2.540 (\emph{0.092}) & 5.080 (\emph{15.480}) & 2.571 (\emph{0.364}) & 5.060 (\emph{15.267}) & 2.904 (\emph{9.403}) & 5.017 (\emph{14.886}) \\
& $\beta_2=-0.125$ & -0.193 (\emph{0.213}) & -0.124 (\emph{0.009}) & -0.238 (\emph{0.836}) & -0.124 (\emph{0.009}) & -0.740 (\emph{20.655}) & -0.130 (\emph{0.011}) \\[1ex]
&$^\diamond \sigma^2_1$ & 0.065  & - & 0.245  & - & 6.364  & - \\
& NAD & 100 & 92 & 100 & 92 & 100 & 94 \\
\hline
\multirow{3}{*}{$n=500$} & $\xi_1=0.1$ & 0.098 (\emph{0.001}) & - & 0.095 (\emph{0.002}) & - & 0.077 (\emph{0.048}) & - \\
& $\beta_1=2.5$ & 2.542 (\emph{0.086}) & 2.633 (\emph{1.840}) & 2.573 (\emph{0.342}) & 4.336 (\emph{6.462}) & 2.852 (\emph{8.649}) & 5.441 (\emph{14.641}) \\
& $\beta_2=-0.125$ & -0.197 (\emph{0.188}) & -0.091 (\emph{0.046}) & -0.242 (\emph{0.742}) & -0.103 (\emph{0.003}) & -0.662 (\emph{18.647}) & -0.095 (\emph{0.008}) \\[1ex]
&$^\diamond \sigma^2_1$ & 0.066  & - & 0.248  & - & 6.126  & - \\
& NAD & 100 & 83 & 100 & 80 & 100 & 89 \\
\hline\hline
\multicolumn{5}{l}{\begin{footnotesize}$\diamond$ results reported as multiplied by $10^2$\end{footnotesize}}
\end{tabular}
\end{table*}
\end{landscape}

\begin{landscape}
\begin{table*}[p]
\centering
\caption{Average posterior means (with \emph{MSE} in brackets) for the parameters of the Lotka-Volterra ODE system; three sample sizes and increasing noise level}\label{lotka_tab}
\begin{tabular}{cl|cc|cc|cc}
\\
\hline\hline
  & \multirow{3}{*}{Parameters} & \multicolumn{6}{c}{Noise level} \\
  \cline{3-8}
 Sample & & \multicolumn{2}{c|}{low ($SNR=13$)} & \multicolumn{2}{c}{medium ($SNR=6.5$)} & \multicolumn{2}{|c}{high ($SNR=1.3$)} \\
 size & & SnM & CollocInfer & SnM & CollocInfer & SnM & CollocInfer \\ 
\hline
\multirow{3}{*}{$n=25$} & $\xi_1=2.0$ & 1.996 (\emph{0.001}) & - & 1.965 (\emph{0.112}) & - & 1.929 (\emph{0.448}) & - \\
& $\beta_1=0.1$ & 0.101 (\emph{0.001}) & 0.095 (\emph{0.001}) & 0.101 (\emph{0.001}) & 0.093 (\emph{0.001}) & 0.091 (\emph{0.001}) & 0.863 (\emph{0.001}) \\
& $\beta_2=-0.2$ & -0.201 (\emph{0.001}) & -0.197 (\emph{0.001}) & -0.197 (\emph{0.001}) & -0.192 (\emph{0.002}) & -0.170 (\emph{0.002}) & -0.183 (\emph{0.008}) \\[1ex]
&$^\diamond \sigma^2_1$ & 0.040  & - & 2.295  & - & 12.234  & - \\
& NAD & 100 & 100 & 100 & 100 & 100 & 100 \\
\hline
\multirow{3}{*}{$n=100$} & $\xi_1=2.0$ & 1.996 (\emph{0.001}) & - & 1.965 (\emph{0.112}) & - & 1.930 (\emph{0.449}) & - \\
& $\beta_1=0.1$ & 0.089 (\emph{0.001}) & 0.063 (\emph{0.001}) & 0.088 (\emph{0.001}) & 0.060 (\emph{0.002}) & 0.086 (\emph{0.001}) & 0.061 (\emph{0.002}) \\
& $\beta_2=-0.2$ & -0.168 (\emph{0.001}) & -0.140 (\emph{0.004}) & -0.167 (\emph{0.001}) & -0.139 (\emph{0.004}) & -0.162 (\emph{0.002}) & -0.138 (\emph{0.004}) \\[1ex]
&$^\diamond \sigma^2_1$ & 3.650  & - & 5.392  & - & 10.507  & - \\
& NAD & 100 & 100 & 100 & 100 & 100 & 98 \\
\hline
\multirow{3}{*}{$n=500$} & $\xi_1=2.0$ & 1.996 (\emph{0.001}) & - & 1.965 (\emph{0.112}) & - & 1.930 (\emph{0.447}) & - \\
& $\beta_1=0.1$ & 0.095 (\emph{0.001}) & 0.080 (\emph{0.001}) & 0.095 (\emph{0.001}) & 0.080 (\emph{0.001}) & 0.094 (\emph{0.002}) & 0.077 (\emph{0.001}) \\
& $\beta_2=-0.2$ & -0.182 (\emph{0.001}) & -0.138 (\emph{0.004}) & -0.182 (\emph{0.001}) & -0.138 (\emph{0.004}) & -0.180 (\emph{0.002}) & -0.138 (\emph{0.004}) \\[1ex]
&$^\diamond \sigma^2_1$ & 1.405  & - & 2.89  & - & 7.407  & - \\
& NAD & 100 & 100 & 100 & 100 & 100 & 99 \\
\hline\hline
\multicolumn{5}{l}{\begin{footnotesize}$\diamond$ results reported as multiplied by $10^1$\end{footnotesize}}
\end{tabular}
\end{table*}
\end{landscape}

\begin{landscape}
\begin{table*}[p]
\centering
\caption{Average posterior means (with \emph{MSE} in brackets) for the parameters of the HIV viral fitness ODE system; three sample sizes and increasing noise level}\label{hiv_tab}
\begin{tabular}{cl|cc|cc|cc}
\\
\hline\hline
  & \multirow{3}{*}{Parameters} & \multicolumn{6}{c}{Noise level} \\
  \cline{3-8}
 Sample & & \multicolumn{2}{c|}{low ($SNR=13$)} & \multicolumn{2}{c}{medium ($SNR=6.5$)} & \multicolumn{2}{|c}{high ($SNR=1.3$)} \\
 size & & SnM & CollocInfer & SnM & CollocInfer & SnM & CollocInfer \\ 
\hline
\multirow{3}{*}{$n=25$} & $\xi_1=60$ & 59.879 (\emph{1.35}) & - & 59.519 (\emph{21.55}) & - & 59.031 (\emph{86.12}) & - \\
& $\beta_1=20$ & 20.808 (\emph{2.79}) & 20.933 (\emph{1.001}) & 20.294 (\emph{21.89}) & 21.142 (\emph{7.130}) & 14.609 (\emph{175.49}) & 20.849 (\emph{8.869}) \\
& $\beta_2=-0.108$ & -0.113 (\emph{0.001}) & - & -0.103 (\emph{0.005}) & - & -0.030 (\emph{0.032}) & - \\
& $^{\dagger}\beta_3=-0.095$ & -0.106 (\emph{0.001}) & -0.106 (\emph{0.003}) & -0.109 (\emph{0.001}) & -0.109 (\emph{0.001}) & -0.159 (\emph{0.001}) & -0.106 (\emph{0.001}) \\[1ex]
&$^\diamond \sigma^2_1$ & 0.169  & - & 2.346  & - & 20.398  & - \\
& NAD & 100 & 84 & 100 & 85 & 100 & 85 \\
\hline
\multirow{3}{*}{$n=100$} & $\xi_1=60$ & 59.882 (\emph{1.18}) & - & 59.539 (\emph{18.82}) & - & 59.079 (\emph{75.26}) & - \\
& $\beta_1=20$ & 21.714 (\emph{3.13}) & 19.700 (\emph{1.936}) & 21.457 (\emph{4.98}) & 19.634 (\emph{2.188}) & 20.281 (\emph{9.94}) & 19.772 (\emph{3.728}) \\
& $\beta_2=-0.108$ & -0.117 (\emph{0.001}) & - & -0.115 (\emph{0.001}) & - & -0.106 (\emph{0.001}) & - \\
& $^{\dagger}\beta_3=-0.095$ & -0.103 (\emph{0.001}) & -0.090 (\emph{0.001}) & -0.103 (\emph{0.001}) & -0.089 (\emph{0.010}) & -0.100 (\emph{0.001}) & -0.090 (\emph{0.009}) \\[1ex]
&$^\diamond \sigma^2_1$ & 0.640  & - & 2.748  & - & 9.473  & - \\
& NAD & 100 & 98 & 100 & 98 & 100 & 97 \\
\hline
\multirow{3}{*}{$n=500$} & $\xi_1=60$ & 59.882 (\emph{1.18}) & - & 59.538 (\emph{18.84}) & - & 59.078 (\emph{75.35}) & - \\
& $\beta_1=20$ & 21.012 (\emph{1.16}) & 19.612 (\emph{1.791}) & 21.040 (\emph{3.17}) & 19.648 (\emph{1.412}) & 20.936 (\emph{8.99}) & 19.499 (\emph{2.437}) \\
& $\beta_2=-0.108$ & -0.114 (\emph{0.001}) & - & -0.113 (\emph{0.001}) & - & -0.112 (\emph{0.001}) & - \\
& $^{\dagger}\beta_3=-0.095$ & -0.100 (\emph{0.001}) & -0.088 (\emph{0.001}) & -0.100 (\emph{0.001}) & -0.089 (\emph{0.001}) & -0.101 (\emph{0.001}) & -0.087 (\emph{0.001}) \\[1ex]
&$^\diamond \sigma^2_1$ & 0.437  & - & 2.312  & - & 8.326  & - \\
& NAD & 100 & 99 & 100 & 99 & 100 & 100 \\
\hline\hline
\multicolumn{5}{l}{\begin{footnotesize}$\dagger$ results reported as multiplied by $10^2$ \end{footnotesize}} \\
\multicolumn{5}{l}{\begin{footnotesize}$\diamond$ results reported as multiplied by $10^{(-1)}$\end{footnotesize}}
\end{tabular}
\end{table*}
\end{landscape}

\appendix
\section{Appendix: Full conditionals for MCMC implementation}
\subsection{Gibbs samplers for the first step (\emph{smooth})}

At iteration $(l+1)$, for each $k=1,\dots,p$, we sample new values for $\boldsymbol{\theta}_k$ from its full conditional distribution

$$
P(\boldsymbol{\theta}^{(l+1)}_k | \dots ) \propto P\!\left( \boldsymbol{y}_{\cdot k} |\, \boldsymbol{\theta}_k, \sigma^{2^{(l)}}_k \right) P\!\left(  \boldsymbol{\theta}_k |\, \lambda^{(l)}_{\theta_k} \right),
$$

which is a multivariate Normal distribution with the following mean vector and variance-covariance matrix 

$$ m_{\theta_k} = V_{\theta_k} \Psi_k \dfrac{\boldsymbol{y}_{\cdot k}}{\sigma^{2^{(l)}}_k} $$
$$ V_{\theta_k} = \biggl[S_{\theta_k}\lambda^{(l)}_{\theta_k} + \dfrac{\Psi_k^{\top} \Psi_k}{\sigma^{2^{(l)}}_k} \biggl]^{-1}$$

where $\Psi_k$ is the basis matrix of the corresponding spline for $\boldsymbol{x}_{\cdot k}$. All noise level components $\sigma^2_k$, except for $\sigma^2_1$, are updated by sampling new values from their full conditional distributions

$$ P( \sigma^{2^{(l+1)}}_k | \dots ) \propto P\!\left( \boldsymbol{y}_{\cdot k} |\, \boldsymbol{\theta}^{(l)}_k, \sigma^2_k \right) P(\sigma^2_k ).$$

These are Inverse Gamma distributions with shape parameter $s_{\sigma^2} = n/2$ and rate parameter equal to

$$ r_{\sigma^2} = \dfrac{\bigl(\boldsymbol{y}_{\cdot k}-\boldsymbol{x}^{(l)}_{\cdot k}\bigl)^{\top} \bigl( \boldsymbol{y}_{\cdot k}-\boldsymbol{x}^{(l)}_{\cdot k} \bigl)}{2}$$

where $\boldsymbol{x}^{(l)}_{\cdot k}$ has to be computed at each iteration given the updated values of $\boldsymbol{\theta}_k$. The full conditional distributions for the penalizing terms $\lambda_{\theta_k}$, with $k=1, \dots, p$, are Gamma distributions with the following shape and rate parameters

$$ s_{\lambda_\theta}=\dfrac{q_k+2}{2}+\alpha_{\theta_k}$$
$$ r_{\lambda_\theta}=\dfrac{\boldsymbol{\theta}^{{(l)}^{\top}}_k S_{\theta_k} \boldsymbol{\theta}^{(l)}_k}{2}+\gamma_{\theta_k},$$

with $q_k$ the number of knots for the corresponding spline and $(\alpha_{\theta_k},\gamma_{\theta_k})$ the hyperparameters governing the prior distribution.

\subsection{Gibbs samplers for the second step (\emph{match})}

At iteration $(l+1)$, we sample new values for $\boldsymbol{\beta}$ from its full conditional distribution

$$ P(\boldsymbol{\beta}^{(l+1)} | \dots ) \propto P\!\left( \boldsymbol{y}^{\star}_{\cdot 1} |\,  \xi^{(l)}_1, \boldsymbol{\beta}, \boldsymbol{\Theta}^{(l)}, \sigma^{2^{(l)}}_1, \right) P\!\left(  \boldsymbol{\beta} |\, \lambda^{(l)}_{\beta} \right) $$

which is a multivariate Normal distribution with the following mean vector and variance-covariance matrix 

$$ m_{\beta} = V_{\beta} \boldsymbol{H}^{(l)} \dfrac{\tilde{\boldsymbol{y}}^{\star}_1}{\sigma^{2^{(l)}}_1} $$
$$ V_{\beta} = \biggl[S_{\beta}\lambda^{(l)}_{\beta} + \dfrac{\boldsymbol{H}^{{(l)}^{\top}} \boldsymbol{H}^{(l)}}{\sigma^{2^{(l)}}_1} \biggl]^{-1}$$

where $\tilde{\boldsymbol{y}}^{\star}_1=\boldsymbol{y}^{\star}_{\cdot 1}-\xi_1^{(l)} \boldsymbol{1}_n$ and $\boldsymbol{H}^{(l)}$ needs to be computed at each iteration by numerically integrating the updated vectors $\boldsymbol{x}^{(l)}_{\cdot k}$. The initial condition $\xi^{(l+1)}_1$ is sampled from a Normal distribution with expected value $m_{\xi}=\boldsymbol{y}^{\star}_{11} - \boldsymbol{H}_1^{(l)}\boldsymbol{\beta}^{(l)}$ and variance $V_{\xi}=\sigma^{2^{(l)}}_1$. As for the penalizing term $\lambda_\beta$ its full conditional distribution is a Gamma distribution with shape and rate parameters equal to

$$ s_{\lambda_\beta}=\dfrac{b}{2}+\alpha_{\beta}$$
$$ r_{\lambda_\beta}=\dfrac{\boldsymbol{\beta}^{{(l)}^{\top}}_k S_{\beta} \boldsymbol{\beta}^{(l)}_k}{2}+\gamma_{\beta},$$

where $b$ is the number of elements in $\boldsymbol{\beta}$ and $(\alpha_{\beta},\gamma_{\beta})$ the hyperparameters governing its prior distribution. Finally, we sample $\sigma^{2^{(l+1)}}_1$ from an Inverse Gamma distribution with parameters

$$ s_{\sigma^2_1}=n$$
\begin{eqnarray*}r_{\sigma^2_1}&=&\dfrac{\bigl(\tilde{\boldsymbol{y}}^{\star}_1-\boldsymbol{H}_1^{(l)}\boldsymbol{\beta}^{(l)} \bigl)^{\top}\bigl(\tilde{\boldsymbol{y}}^{\star}_1-\boldsymbol{H}_1^{(l)}\boldsymbol{\beta}^{(l)} \bigl)}{2}+\\
&+&\dfrac{\bigl(\boldsymbol{y}_{\cdot 1}-\boldsymbol{x}^{(l)}_{\cdot 1}\bigl)^{\top} \bigl( \boldsymbol{y}_{\cdot 1}-\boldsymbol{x}^{(l)}_{\cdot 1} \bigl)}{2}.
\end{eqnarray*}

\end{document}